\documentclass[%
 reprint,
 amsmath,amssymb,
 aps,
prb,
]{revtex4-2}

\usepackage{graphicx}
\usepackage{dcolumn}
\usepackage{bm,bbm,mathrsfs}
\usepackage{xcolor}
\usepackage[colorlinks=true,linkcolor=blue,citecolor=blue]{hyperref}
\newcommand{\ket}[1]{|#1\rangle}
\newcommand{\bra}[1]{\langle#1|}

\newcommand{\inp}[2]{\langle#1|#2\rangle}

\newcolumntype{L}[1]{>{\raggedright\arraybackslash}p{#1}}
\newcolumntype{C}[1]{>{\centering\arraybackslash}p{#1}}
\newcolumntype{R}[1]{>{\raggedleft\arraybackslash}p{#1}}

			\newcommand{\e}[1]{\begin{align}{#1}\end{align}}	
			\newcommand{\bkp}{\bm{k}_{\perp}}
		
		\newcommand{\f}[2]{\frac{#1}{#2}}

		\newcommand{\p}[2]{\frac{\partial #1}{\partial #2}}

		\newcommand{\la}[1]{\label{#1}}

		\newcommand{\q}[1]{Eq.\ (\ref{#1})}
		\newcommand{\qq}[2]{Eqs.\ (\ref{#1})-(\ref{#2})}
		\newcommand{\s}[1]{Sec.\ \ref{#1}}
		\newcommand{\fig}[1]{Fig.\ \ref{#1}}

		\newcommand{\ocite}[1]{Ref.\ \onlinecite{#1}}

		\newcommand{\ri}{\rightarrow}

		\newcommand{\sgn}{\text{sgn}}

		\newcommand{\eq}{=&\;}

		\newcommand{\Z}{\mathbb{Z}}

\newcommand\as{\;\;\;\;}

\newcommand{\bk}{\boldsymbol{k}}

\newcommand{\bn}{\boldsymbol{n}}

\newcommand{\bM}{\boldsymbol{M}}

\newcommand{\cali}{{\cal I}}

\newcommand{\calp}{{\cal P}}

\newcommand{\tr}{\text{Tr}}

\newcommand{\noi}[1]{\noindent (#1)}

\newcommand{\half}{\frac{1}{2} }

\newcommand{\lin}{\notag \\}

\newcommand{\bpm}{\begin{pmatrix}}
\newcommand{\epm}{\end{pmatrix}}

\newcommand{\bal}{\begin{align}}

\newcommand{\dg}[1]{#1^{\scriptstyle{\dagger}}}

        \definecolor{AAcolor}{rgb}{0.7,0.1,0.4}

\begin{document}

\preprint{APS/123-QED}

\title{ Quantized surface magnetism and higher-order topology:
\\
Application to the Hopf insulator}

 \author{Penghao Zhu} \affiliation{Department of Physics and Institute for Condensed Matter Theory, University of Illinois at Urbana-Champaign, Urbana, Illinois 61801, USA}
 \author{Taylor L. Hughes} \affiliation{Department of Physics and Institute for Condensed Matter Theory, University of Illinois at Urbana-Champaign, Urbana, Illinois 61801, USA}
 \author{A. Alexandradinata} \altaffiliation{Corresponding author: aalexan7@illinois.edu} \affiliation{Department of Physics and Institute for Condensed Matter Theory, University of Illinois at Urbana-Champaign, Urbana, Illinois 61801, USA}




\date{\today}

\begin{abstract}
We identify topological aspects of the subextensive magnetic moment contributed by the surfaces of a three-dimensional crystallite -- assumed to be insulating in the bulk  as well as on all surface facets, with trivial Chern invariants in the bulk.
The geometric component of this subextensive moment is given by its derivative with respect to the chemical potential, at zero temperature and zero field, per unit surface area, and hence corresponds to the surface \textit{magnetic compressibility}. The sum of the surface compressibilities  contributed by two opposite facets of a cube-shaped crystallite is quantized to an integer multiple of the fundamental constant $e/h c$; this integer is in one-to-one correspondence with the net chirality of hinge modes on the surface of the crystallite, manifesting a link with  higher-order topology. The contribution by a single facet to the magnetic compressibility need \textit{not} be quantized to integers; however, symmetry and/or Hilbert-space constraints can fix the single-facet compressibility to \textit{half}-integer multiples of $e/hc$, as will be exemplified by the Hopf insulator. \end{abstract}
\maketitle

Disentangling bulk and boundary effects in orbital magnetism \cite{thonhauser2005orbital,ceresoli2006orbital,shi2007quantum} has long been a subtle affair. Boundary currents contribute extensively to the orbital magnetic moment, yet  the boundary current  cannot be completely attributed to single-particle states localized to the boundary \cite{thonhauser2011theory, bianco2016orbital}. The recent formulation of a local marker for the orbital magnetization allows one, for the first time, to unambiguously distinguish boundary contributions to the orbital magnetic moment from bulk contributions \cite{bianco2013orbital}. For band insulators, it was  subsequently realized that the boundary contribution can be \textit{extensive} only if the insulator has a nonzero Hall conductivity\cite{bianco2016orbital}.


What is missed in previous works is the possibility of boundary contributions to the magnetic moment which scale \textit{subextensively} with the area of the surface in the thermodynamic limit.  This work aims to identify topological aspects of the subextensive magnetic moment contributed by surfaces of a three-dimensional crystallite -- with \textit{trivial} Chern invariants in the bulk. Such aspects arise when there is a spectral gap to excitations not just in the bulk, but also on all surface facets, as is assumed throughout this work. 

\begin{figure}[h]
\centering
\includegraphics[width=1\columnwidth]{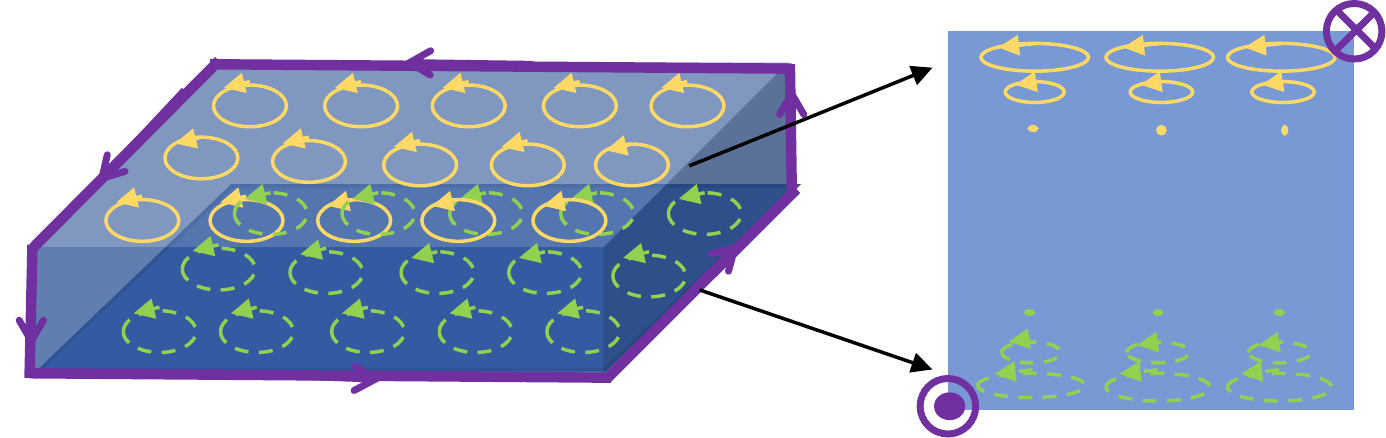}
\caption{Schematic illustration of a cubic slab with surface magnetism (indicated by current loops) and hinge modes (purple arrows). On the right, a cross section of the left slab illustrates the exponential decay of the surface magnetization. }
\label{fig:illustration}
\end{figure}

For illustration, consider the magnetic moment contributed by the top and bottom facets of a cubic slab depicted in Fig. \ref{fig:illustration}. The geometric component of this subextensive moment, as we will show, is given by its derivative with respect to the chemical potential, at zero temperature and zero field, per unit surface area.  This quantity, henceforth referred to as the magnetic compressibility, is known to be quantized to integer multiples of the fundamental constant $e/hc$ for strictly two-dimensional, {bulk insulators}, and the integer invariant has been identified with the net chirality of the edge modes \cite{ceresoli2006orbital, bianco2013orbital,schulz2013orbital,bianco2016orbital}. Generalizing this statement to a three-dimensional cubic slab, the analogous integer invariant becomes the net chirality of hinge modes localized to the one-dimensional intersections of the four side facets, as illustrated in Fig. \ref{fig:ampere}.  This means that a three-dimensional insulator with a nontrivial magnetic compressibility is also a higher-order topological insulator \cite{Langbehn2017Reflection,benalcazar2018prb,schindler2018higher,schindler2018bismuth,Kooi2018inversion,PhysRevB.97.205135,PhysRevX.9.011012}, as we will establish.

This is one of several links between surface magnetism and higher-order topology that we will explore in \s{sec: surfaceM_sahc}. Unlike strictly two-dimensional systems, the contribution of a single facet to the magnetic compressibility need \textit{not} be quantized to integers. However, we find that certain symmetries, and/or Hilbert-space constraints, can fix the single-facet compressibility to \textit{half}-integer multiples of $e/hc$. Our main case study [cf.\ \s{sec: hopf}] for this half-integer quantization is the Hopf insulator \cite{PhysRevLett.101.186805, PhysRevB.88.201105}, a three-dimensional magnetic topological insulator inspired by the Hopf map in differential topology \cite{hopf1964abbildungen}. 


Beyond the Hopf insulator, we also consider a similar (but strictly distinct) class of insulators with zero bulk magnetization at zero field, but with nontrivially-quantized, single-facet magnetic compressibility. Such a phase of matter does not fit into standard classifications under the family tree of magnetism \cite{mattis2006theory} (paramagnetism, ferromagnetism, ferrimagnetism, etc.), and we refer to it as a  \textit{topomagnetic insulator}.
In \s{sec:topomagnet}, we show that some (but not all) higher-order topological insulators known in the literature\cite{PhysRevLett.98.106803,qi2008topological,sitte2012topological,schindler2018higher} are topomagnetic insulators. The concluding \s{sec: conclusions} contains a partial summary of our results and suggestions on how the subextensive magnetic moment can be measured.

\section{Faceted orbital magnetization and higher-order topology}
\label{sec: surfaceM_sahc}


After a brief review of the local marker for orbital magnetization in  Sec.  \ref{sec: localM}, we present a formula for the layer-resolved magnetization in \s{sec: surfM}, which is used to isolate a single facet's contribution to the  orbital magnetic moment. The associated compressibility of this moment is shown to relate to the geometric component of the surface anomalous Hall conductivity  in Sec. \ref{sec: compressibility_hinge}. We also derive therein the correspondence between compressibility of facets and chiral hinge modes, by arguments involving Amperean loops and the Streda formula.

\subsection{Review of local marker for orbital magnetization}
\label{sec: localM}

For an insulating crystallite in contact with an electron reservoir (at chemical potential $\mu$), the zero-temperature magnetization  is  related  to the grand thermodynamic potential $\Omega$ as
\e{M= -\f1{V}\p{\Omega}{B}\bigg|_{\mu}= -\f1{V}\p{E}{B}+\mu\p{n}{B}:=m_1+m_2,\la{mgrand}}
where $n=\f{N}{V}$ is the electron density, $E$ is the internal energy, and $B$ is the external magnetic field. The $m_1$ term  contains an orbital contribution originating from the minimal coupling to the magnetic field, as well as a spin contribution originating from the Zeeman coupling to the intrinsic spin magnetic moment. We refer to the orbital component of $m_1$ plus the entirety of $m_2$ as the orbital magnetization $m_{orb}$.

We shall primarily be concerned with the zero-$B$ orbital magnetization of band insulators with a spectral gap in both the bulk and all surface facets of a three-dimensional crystallite. Since we are not considering a fully periodic system with translation symmetry it is important that we introduce a method to calculate the magnetization in position-space. Fortunately, it was already shown in Ref \onlinecite{bianco2013orbital} that $m_{orb}$ can be expressed as an integral over continuous position-space of a local marker $\mathfrak{m}$ for orbital magnetization. For convenience, we present here the tight-binding analog of their formula: 
\begin{equation}
\label{eq: mag_bulk}
\begin{aligned}
&M_{\gamma}=\frac{1}{V}\sum_{\bm{R}}\mathfrak{m}_{\gamma}(\bm{R}),
\\
\mathfrak{m}_{\gamma}(\bm{R}) &=\frac{e}{2\hbar c}\epsilon_{\alpha\beta\gamma} \operatorname{Im}
\sum_{a}\bra{\bm{R},a}\bar{P} \hat{r}_{\alpha}\bar{Q} \bar{H} \bar{Q} \hat{r}_{\beta} \bar{P}
\\
&\quad -\bar{Q}\hat{r}_{\alpha}\bar{P}(\bar{H}-2\mu)\bar{P}\hat{r}_{\beta}\bar{Q}\ket{\bm{R},a},
\end{aligned}
\end{equation}
where $\bar{H}$ is the Hamiltonian, and $\bar{P}$ and $\bar{Q}$ are the projectors onto the occupied and unoccupied subspaces of $\bar{H}$, respectively. $\bar{H},\bar{P}$ and $\bar{Q}$ act in a Hilbert space spanned by the orthonormal basis vectors $\{\ket{\bm{R},a}\}_{\bm{R},a}$, where $\ket{\bm{R},a}$ is a ket state of an electron localized (as a Wannier function) to a Bravais-lattice vector $\bm{R}=(R_{x},R_{y},R_{z})$, and the index $a$  labels the basis vectors of the reduced Hilbert space of one primitive unit cell. The action of the position operator $\hat{r}_{\alpha}$ on these states satisfies $\hat{r}_{\alpha}\ket{\bm{R},a}=R_{\alpha}\ket{\bm{R},a}$.  By assumption, an energy gap separates the occupied and unoccupied subspaces of $\bar{H}$, so that the matrix elements of $\bar{P}$ decay exponentially with increasing position-space separation \cite{PhysRev.135.A685,PhysRevB.74.235111}. This gives an exponentially sharp distinction between the local marker  $\mathfrak{m}_{\gamma}(\bm{R})$ evaluated on the boundary versus in the bulk \cite{bianco2013orbital}.  Using this formalism we immediately observe that the $\mu$-dependent term in \q{eq: mag_bulk} is geometric, i.e., it is simply proportional to the local Chern marker \cite{PhysRevB.84.241106}. Alternatively, this proportionality can be derived by applying the Streda formula \cite{streda1982theory,RevModPhys.82.1959} to  $m_2= \mu \partial n/\partial B$.



\subsection{Faceted orbital magnetization}
\label{sec: surfM}

To model a facet on a three-dimensional crystallite, let us consider a half-infinite slab with surface normal vector $+\hat{z}$. The orbital magnetic moment of the slab can be decomposed into a sum over layers indexed by $l=1,2,\ldots$, where $l=1$ lies closest to the surface termination. By summing the local marker over each layer, we show in Appendix \ref{app: deri_maghf2} that the  orbital magnetization (contributed by each layer) is expressible as an integral over the reduced Brillouin zone ($rBZ \ni \bk_{\perp}=(k_x,k_y) $):
\begin{equation}
\label{eq: Mghf2}
\begin{aligned}
&M_{z}(l):=\frac{e}{c\hbar} \operatorname{Im}\int_{rBZ} \frac{d \mathbf{k}_{\perp}}{(2 \pi)^{2}}\bigg[(g_{\mathbf{k}_{\perp},xy}(l)
\\
&+h_{\mathbf{k}_{\perp},xy}(l))-2\mu f_{\mathbf{k}_{\perp},xy}(l))\bigg],
\end{aligned}
\end{equation}
\begin{equation}
\label{eq: g&h2}
\begin{array}{c}g_{\mathbf{k}_{\perp}, xy}(l)=\operatorname{Tr}_{\mathrm{cell},z}\left[\mathcal{P}_{l}\partial_{k_x} \widetilde{P}\widetilde{Q} \widetilde{H}\widetilde{Q}\partial_{k_y} \widetilde{P}\right],
\\ 
h_{\mathbf{k}_{\perp}, xy}(l)=\operatorname{Tr}_{\mathrm{cell},z}\left[\mathcal{P}_{l}\widetilde{Q}\partial_{k_y} \widetilde{P}\widetilde{H}\partial_{k_x} \widetilde{P}\widetilde{Q} \right],
\\
f_{\mathbf{k}_{\perp}, xy}(l)=\operatorname{Tr}_{\mathrm{cell},z}\left[\mathcal{P}_{l}\widetilde{Q}\partial_{k_y} \widetilde{P}\partial_{k_x} \widetilde{P}\widetilde{Q}\right],
\end{array}
\end{equation}
where $g,h,f$ correspond, in order, to the three terms in \q{eq: mag_bulk}.
The quantity $\widetilde{P}(\mathbf{k}_{\perp})$ ($\widetilde{Q}(\mathbf{k}_{\perp})$) is the projector to the occupied (unoccupied) subspace of the Hamiltonian  $\widetilde{H}(\mathbf{k}_{\perp})$; note the arguments have been suppressed in the above equations above. In a slab geometry, these operators have matrix elements labelled as, for example, $\widetilde{H}(\mathbf{k}_{\perp})_{R_z R_{z^{\prime}},ab}$, and  $``\operatorname{Tr}_{\mathrm{cell},z}$" is the trace with respect to the composite index $(R_z,a)$. Finally, the operator $\mathcal{P}_{l}(\mathbf{k}_{\perp})$ is the projector onto the $R_{z}=l$ layer in the $\mathbf{k}_{\perp}$ momentum sector.  In direct correspondence with the local Chern marker \cite{PhysRevB.84.241106}, the imaginary part of $f$ in  \q{eq: Mghf2} is the layer-resolved Berry curvature \cite{essin2009magnetoelectric}, and its integral over the $rBZ$ gives the contribution (of each layer) to the Hall conductivity:
\begin{equation}
C_{+\hat{z}}(l)=\frac{1}{2\pi}\int_{rBZ} d\bm{k}_{\perp}\operatorname{Im}f_{\mathbf{k}_{\perp},xy}(l).\la{layerHallcond}
\end{equation}

Now we can define the \emph{faceted orbital magnetization} as the orbital magnetization contributed by the slab surface (or more generally a facet of a three-dimensional crystallite). This quantity is obtained by summing the layer-resolved magnetization in the vicinity of the surface, and subtracting the bulk contribution:
\begin{equation}
\label{eq: surfaceM}
M_{+\hat{z}}=\sum_{l=1}^{bulk} \bigg(M_{z}(l)-M_{z,bulk}\bigg),
\end{equation}
with $M_{z,bulk}$ defined as the layer-resolved magnetization in the bulk region; in a slab geometry, $M_{z,bulk}$ is the asymptotic value of $M_{z}(l)$ for large $l$.  By $\sum_{l=1}^{bulk}$, we mean that the summation should be carried out into the bulk region. The exponential decay of  $\tilde{P}_{R_zR_z';ab}$ in $|R_z-R_z'|$ guarantees that the sum reaches its asymptotic value exponentially. The magnetic moment contributed by a facet of area $A$ is then $M_{+\hat{z}}A$ -- this form manifests that the faceted magnetic moment is subextensive.



\subsection{Faceted magnetic compressibility and chiral hinge modes}
\label{sec: compressibility_hinge}

Differentiating \q{eq: surfaceM} with respect to the chemical potential, we obtain the magnetic compressibility contributed by a single facet (in short, the \textit{faceted compressibility}):
\begin{equation}
\label{eq: dmdmu2}
\frac{d M_{+\hat{z}}}{d\mu}=-\frac{e}{hc}\sum_{l=1}^{bulk}C_{+\hat{z}}(l).
\end{equation}
The above equation can be derived straightforwardly from \qq{eq: Mghf2}{layerHallcond}. Note that the derivative of $M_{z,bulk}$ vanishes by our initial assumption of the triviality of the bulk Chern invariants. Hence, we can then identify
$\sum_{l=1}^{bulk}C_{+\hat{z}}(l) e^2/h$ as the geometric component of the surface anomalous Hall conductivity (in short, the geometric SAHC, which we denote by $\sigma_{+\hat{z}}^{g}$).\cite{essin2009magnetoelectric,rauch2018geometric}. More generally, for any facet with normal vector $\hat{\bm{n}}$,
\begin{equation}
\label{eq: dmdmu}
\frac{d M_{\hat{\bm{n}}}}{d\mu}=-\frac{1}{ec}\sigma_{\hat{\bm{n}}}^{g},
\end{equation}
where $\sigma_{\hat{\bm{n}}}^{g}$ is the geometric SAHC on the surface facet with normal vector $\hat{\bm{n}}$.
Because the spin contribution to the magnetization does not depend on the chemical potential [cf.\ \s{sec: localM}], the above equation represents the complete compressibility contributed by a single facet, and not just its orbital component.

\begin{figure}[h]
\centering
\includegraphics[width=0.8\columnwidth]{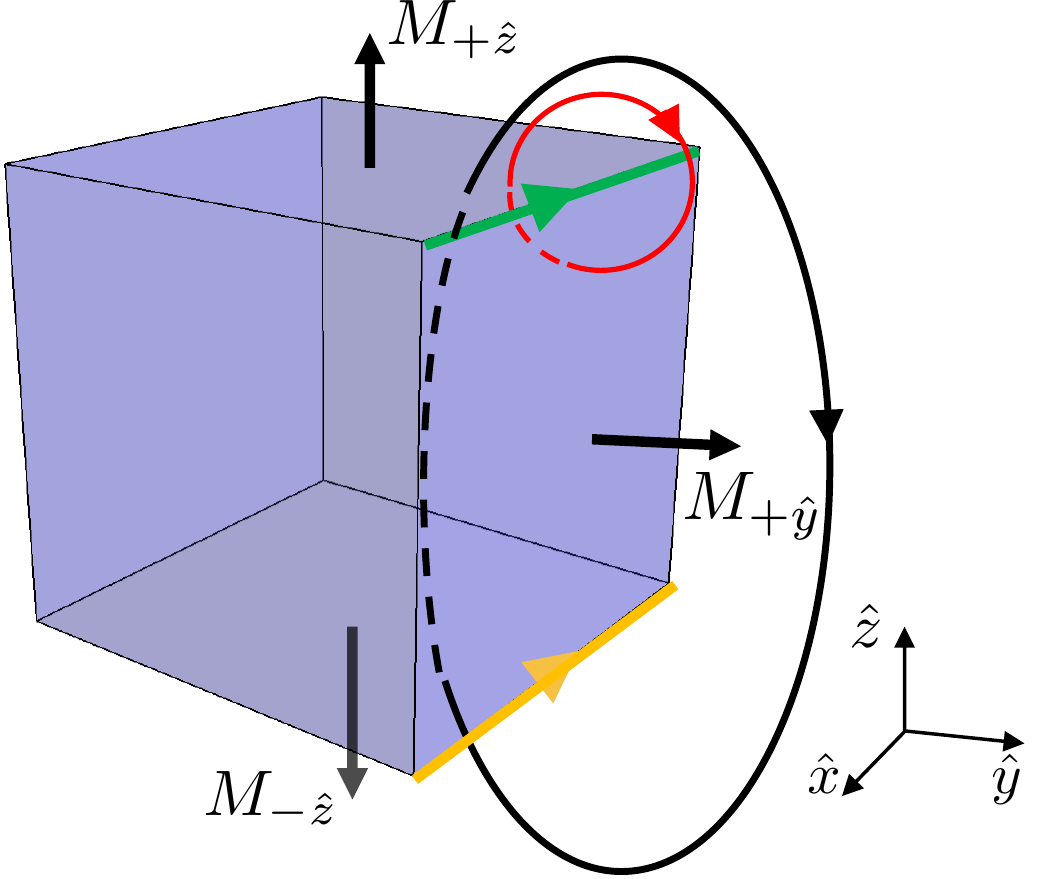}
\caption{Geometry of facets (blue), hinges (green and orange), and Amperean loops (red and black). The arrows on facets indicate the direction of the faceted magnetic moment, assuming $M_{\hat{\bm{n}}}>0$ for each of $\hat{\bm{n}}\in \{+\hat{z},+\hat{y},-\hat{z}\}$.}
\label{fig:ampere}
\end{figure}

Let us emphasize a distinction between the faceted magnetization and the magnetization of a strictly two-dimensional system $M_{2D}$.  The latter satisfies $dM_{2D}/d\mu=-(e/\hbar c)C$ where $C$ is the integer-valued Chern number \cite{bianco2013orbital,ceresoli2006orbital}, which remains quantized even in disordered systems\cite{schulz2013orbital}. In contrast, $M_{\hat{\bm{n}}}$, as defined on a surface facet of a three-dimensional crystallite, satisfies Eq. \eqref{eq: dmdmu2}. Here the quantity $\sum_{l=1}^{bulk}C_{\hat{\bm{n}}}(l)$  is not necessarily fixed to discrete values. However, we will now show that the difference of $\sum_{l=1}^{bulk}C_{\hat{\bm{n}}}(l)$ on adjacent facets must be quantized to an integer, with this integer equal to (minus) the net chirality of the hinge modes at the interface of the two facets.

Let us derive this equality from the classical electromagnetic relation between magnetization and current: $\oint \mathbf{M}\cdot d\mathbf{l}=I/c$ \cite{jackson2007classical}, with $\bM$ the total magnetization defined per unit volume, as in \q{eq: mag_bulk}. We construct an Amperean loop to encircle a hinge between two facets, as illustrated by the red loop in Fig. \ref{fig:ampere}. By assumption, the local Chern marker in the bulk is trivial, hence $d\bm{M}/d\mu$ receives contributions only from the faceted magnetization: $M_{+\hat{z}}\delta(z-z_0)\hat{z}$ for the top facet (colored blue in Fig. \ref{fig:ampere}), and $M_{+\hat{y}}\delta(y-y_0)\hat{y}$ for the right facet (also colored blue). This means that
\begin{equation}
\label{eq: dIdc}
\begin{aligned}
\frac{dM_{+\hat{z}}}{d\mu}-\frac{dM_{+\hat{y}}}{d\mu}=\f1{c}\frac{dI}{d\mu},
\end{aligned}
\end{equation}
for a current $I$ flowing through the Amperean loop.
By assumption, there is an energy gap to excitations for both bulk and surface states. Thus,  a variation of $\mu$ within this gap can only change the filling of hinge states localized to the interface colored green in \fig{fig:ampere}, and $dI/d\mu$ is completely determined by the hinge state current. Each hinge state with wave number $k$ and velocity $v(k)$ contributes $-ev(k)/L$ to the current $I$. Let  $s$ be a label for each hinge mode crossing the Fermi level with  velocity $v_F^{(s)}$ and  wave number $k_F^{(s)}$. For an infinitesimal increase $\delta\mu$, the resultant change in current is 
\e{\delta I=-\f{e}{L} \sum_{s} \f{L|\delta k_F^{(s)}|}{2\pi} v_F^{(s)}, }
where $\delta k_F^{(s)}=\delta \mu/v^{(s)}\hbar$ is the change in Fermi wave number for mode $s$. It follows that $\delta I/\delta \mu =-(e/h) \sum_s \sgn[v_F^{(s)}]$, hence
\e{\frac{d M_{+\hat{z}}}{d\mu}-\frac{d M_{+\hat{y}}}{d\mu}=-\f{e}{hc}\sum_s \sgn[v_F^{(s)}], \la{chirality} }
which proves our claim. We remark that \q{chirality} is not unexpected given that the faceted compressibility is related to the geometric SAHC through Eq.\eqref{eq: dmdmu}; it is well-known that the differential geometric SAHC between two adjacent facets determines the chirality of states localized at their interface \cite{sitte2012topological,PhysRevB.98.245117,PhysRevB.102.035166}. Our present derivation, however, emphasizes the heretofore unexplored connection between chiral hinge states and the faceted compressibility.  


By a simple generalization of the above argument to the black Amperean loop in Fig. \ref{fig:ampere}, we derive 
\e{\frac{d M_{+\hat{z}}}{d\mu}-\frac{d M_{-\hat{z}}}{d\mu}=-\f{e}{hc}\bigg(\sum_s \sgn[v_F^{(s)}]+\sum_{s'} \sgn[v_F^{(s')}]\bigg),\la{simplegen}}
where $M_{-\hat{z}}$ is the magnetization of the bottom facet, and $s'$ is an index for hinge modes localized to the bottom hinge colored orange in  \fig{fig:ampere}. The relative minus sign on the left-hand side of \q{simplegen} is because the faceted magnetization is defined to point from the interior to the exterior. 

Let us present an alternative and instructive derivation of  \q{simplegen}. 
Because the bulk magnetization has vanishing compressibility (since bulk Chern numbers vanish), the right-hand side of \q{simplegen} may be identified with the compressibility of the total magnetization: $M_z=\sum_l M_z(l)$, where the sum is over all layers in the slab; $M_z$ is defined with respect to a fixed Euclidean coordinate system indicated on the right-bottom corner of \fig{fig:ampere}. Applying the thermodynamic relation [in \q{mgrand}] to the case of a magnetization defined \emph{per unit area}, \begin{equation}
\label{eq: streda}
\p{M_z}{\mu}=\frac{1}{A_{xy}}\frac{\partial N}{\partial B}=\frac{e}{hc}C_{\mathrm{slab},z},
\end{equation}
where $N$ is the total electron number and $A_{xy}$ is the area of the slab. The second equality above relates  $\partial N/\partial B$ (at fixed $\mu$ and zero temperature) to the first Chern number ($C_{\mathrm{slab},z}$) of the slab, in accordance with Streda's formula \cite{streda1982theory}. By the bulk-boundary correspondence for Chern insulators, there must be chiral modes localized to the `edge' of the slab. In this case, the `edge' is a macroscopically large side-surface facet. However, we have assumed that any surface states are gapped, and thus the chiral `edge' modes are identified with hinge modes localized to either hinge.

The formulation of a faceted magnetization [cf.\ Eqs. \eqref{eq: Mghf2}, \eqref{eq: g&h2}, \eqref{eq: surfaceM}], and the relation -- between the faceted compressibility and the geometric SAHC [Eqs. \eqref{eq: dmdmu}] and chiral hinge states [Eqs. \eqref{chirality},\eqref{simplegen}] -- are the main results of this section. With these formal preparations out of the way, we proceed to an illustrative example.

\section{Case study of the Hopf insulator}
\label{sec: hopf}

The Hopf insulator is a three-dimensional bulk insulator with a two-band, bulk Hamiltonian $H(\bk)$ that cannot be continuously deformed to a trivial $\bk$-independent Hamiltonian while the bulk energy gap and bulk translation symmetry are preserved \cite{PhysRevLett.101.186805,PhysRevB.88.201105,alexandradinata2019actually}. This topological obstruction exists despite the Hopf insulator having trivial first Chern class -- meaning the Chern number of any two-dimensional slice in the three-dimensional Brillouin zone vanishes. (We leave to future investigations the related notion of Hopf-Chern insulators  \cite{kennedy2016topological}, which have a nontrivial first Chern class in the bulk.) Instead of a bulk Chern invariant, the bulk topology of the Hopf insulator is diagnosed by the \textit{Hopf invariant}, which is equivalent to a BZ-integral of the Abelian Chern-Simons three-form \cite{PhysRevLett.101.186805}:
\begin{equation}
\label{eq:hopfnumber}
\chi=-\frac{1}{4\pi^{2}}\int_{BZ}d^{3}k\bm{A}(\bm{k})\cdot\bm{\mathcal{F}}(\bm{k}),
\end{equation}
where $\bm{A}(\bm{k})=i\inp{u}{\nabla_{\bm{k}}u}$ is the Berry connection of the valence band, and $\bm{\mathcal{F}}(\bm{k})=\nabla \times \bm{A}(\bm{k})$ is the associated Berry curvature.  This Chern-Simons formula is gauge-invariant if the valence band has unit rank, and is furthermore quantized to integer values if the conduction band \textit{also} has unit rank; both conditions are automatically satisfied because the model we use for $H(\bk)$ is a sum of Pauli matrices. (The \textit{rank} of a band defined over the BZ or rBZ is defined as the number of independent wave functions that span the band at each $\bm{k}\in BZ$, resp.\ $\bk_{\perp}\in rBZ$.)

Being an axion insulator with quantized magneto-electric polarizability \cite{alexandradinata2019actually}, the Hopf insulator is a prime candidate for the quantized magnetic compressibility  introduced in \s{sec: localM}. Additionally, the Hopf insulator presents a unique opportunity to extend the Chern-Simons/axion theory of magneto-electric polarizability to the case where the Chern-Simons action is gauge-invariant, unlike other known axion insulators \cite{essin2009magnetoelectric,essin2010orbital,malashevich2010theory,coh2011chern,thonhauser2011theory,sitte2012topological,PhysRevB.84.075119,PhysRevLett.98.106803,qi2008topological,vanderbilt_2018,rauch2018geometric}.

Let us summarize the main results of our case study:\\

\noi{i} The faceted magnetic compressibility of the Hopf insulator is a half-integer multiple of $e/hc$:
\begin{equation}
\label{eq: dmdmu_hopf}
\frac{d M_{\hat{\bm{n}}}}{d\mu}=\frac{e}{hc}\left(\frac{\chi}{2}-C_{v}(\hat{\bm{n}})\right).
\end{equation}
The half-integer includes a contribution from the bulk integer invariant $\chi \in \Z$, and also  a surface-dependent contribution from the integer invariant $C_{v}(\hat{\bm{n}})\in \Z$. Roughly speaking, $C_{v}$ is the first Chern number of the occupied bands localized to a surface facet;  the precise formulation of this quantity requires a  Bloch-Wannier representation that is elaborated in \s{sec: classification}.\\

\noi{ii} \q{eq: dmdmu_hopf} is derived by relating [cf.\ \q{eq: dmdmu}] the facet compressibility to the geometric component of the surface anomalous Hall conductivity (SAHC); the latter quantity is related to the aforementioned topological invariants by
\e{\sigma^{g}_{\hat{\bm{n}}}=\left(-\frac{\chi}{2}+C_{v}(\hat{\bm{n}})\right)\frac{e^2}{h},\la{geometricsahchopf}}
as we prove in  \s{sec:sahc_hopf}.\\


\noi{iii} From \q{geometricsahchopf}, we argue that if two neighboring facets differ in their values for $C_{v}(\hat{\bm{n}})$, then there must be chiral hinge mode(s) localized to the interface of the two facets -- this follows from a generalized Laughlin argument (for surface facets) that is described in \s{sec:sahctrivial}. Furthermore, we show for the Hopf insulator that the existence of hinge modes is guaranteed by a symmetry that combines particle-hole conjugation with spatial inversion, as demonstrated in \s{sec: chiral_difference}.\\


\noi{iv} Finally in  \s{sec:numericcompress}, we numerically simulate the quantization of \q{eq: dmdmu_hopf} for a model Hamiltonian.


\subsection{Formulation of a Chern number for states localized to a facet}
\label{sec: classification}

\begin{figure}[h]
\centering
\includegraphics[width=1\columnwidth]{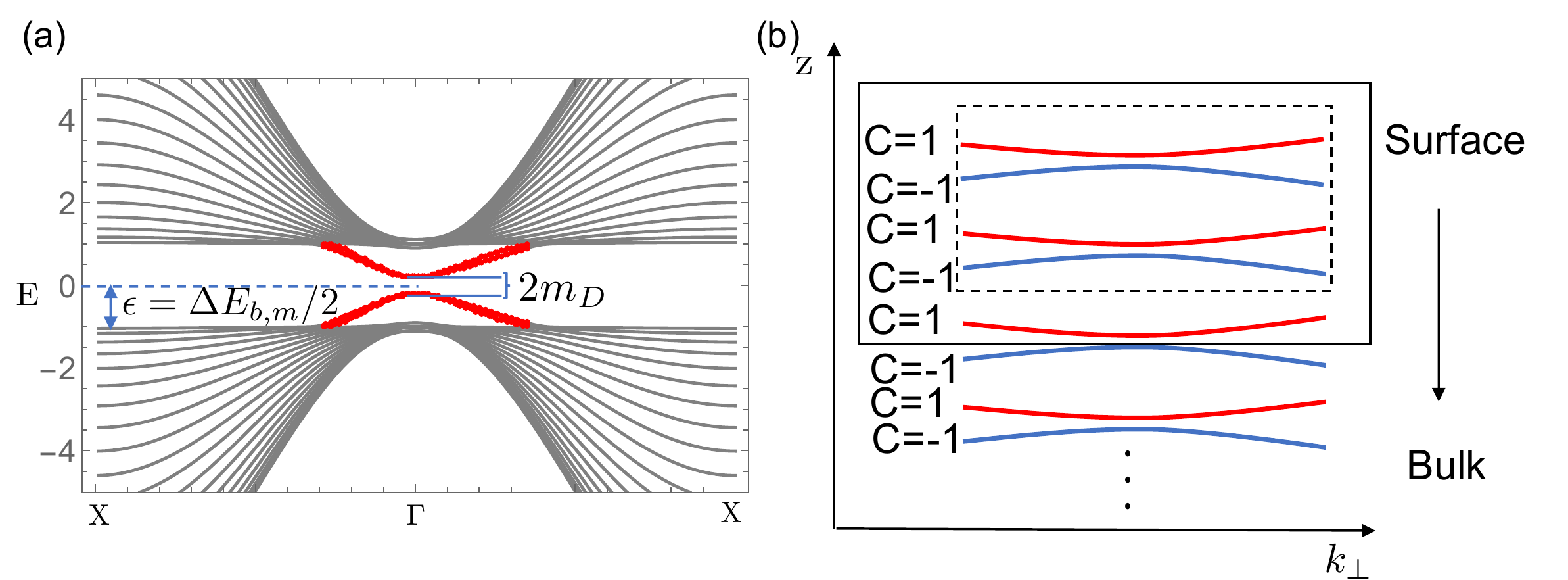}
\caption{(a) Partial surface bands in the open-boundary spectrum of the model Hamiltonian in \q{eq: MRW},  with $u=2$. The states colored red  are localized to surface facets. $\Delta E_{b,m}$ is minimal energy gap in the bulk. $\epsilon$ is set to be the natural energy scale in this paper. $m_{D}$ is the surface Dirac mass. (b) Schematic illustration of Bloch-Wannier bands for a superlattice model described in \s{sec:csv}. }
\label{fig:schematic}
\end{figure}

We have proposed, roughly speaking, that the integer invariant $C_{v}$ in \q{eq: dmdmu_hopf} is the Chern number for occupied states localized to a facet. To have a properly defined Chern number (as an integral of the Berry curvature over the reduced Brillouin zone $rBZ$), it is necessary to identify an entire band of occupied states that is localized to the facet, with a projector that is continuously defined over the \emph{entire} $rBZ$. This does not generally hold when one evaluates \textit{energy eigenstates} of topological band systems with a surface termination. The more likely  scenario is that of a `{partial surface band}', whose energy spectrum is attached to a bulk energy band in some region of the $rBZ,$ such that the states localized to a facet (in position space) are confined to a subregion $\mathfrak{D}$ of the $rBZ$ (in  momentum space), as exemplified in Fig. \ref{fig:schematic}(a). 
In general, the integral of the Berry curvature over $\mathfrak{D}$ is not quantized to integer values.

To obtain an integer-valued Chern invariant for surface-localized states in the more general scenario of a partial surface band, one has to relax our previous assumption that said states are energy eigenstates of the single-particle Hamiltonian. Instead, we propose to define the Chern invariant $C_{v}$ for surface-localized states that are eigenstates of the projected position operator on a half-infinite geometry, as  explained in Sec \ref{sec:csv}. (Note that eigenstates of the projected position operator are simply linear combinations of the occupied energy eigenstates.) Then, in Sec \ref{sec: csvtopo}, we describe the conditions that make $C_{v}$ a well-defined topological invariant. One implication is that if  a three-dimensional, bounded crystallite  is insulating in the bulk and on all surface facets, and has a bulk Pauli-matrix Hamiltonian in the trivial first Chern class, then bulk translational symmetry \textit{alone} gives a topological classification specified the bulk invariant $\chi$ [cf.\ \q{eq:hopfnumber}] and the set of $C_{v}$ for all facets.

\subsubsection{Chern invariant  $C_{v}$ from the projected position operator}\la{sec:csv}

Given a bulk insulator with an occupied subspace of states (represented by the projector $P$) and an unoccupied subspace (with a projector $Q$), we consider the projected position operators $P\hat{Z}P$ and $Q\hat{Z}Q$, with $\hat{Z}$  the position operator in the $z$ direction. Throughout this subsection we impose periodic boundary conditions in the $x$ and $y$ directions.

While the eventual goal of this section is to define the Chern invariant $C_{v}$ through the projected position operator defined on a half-infinite slab geometry (with $\hat{Z}$ taking only positive values), it is useful to review some properties of $P\hat{Z}P$ on an infinite geometry (with $\hat{Z}$ taking also negative values). These properties hold equally well for $Q\hat{Z}Q$, with the semantic replacement $P\rightarrow Q$. Since translations perpendicular to $z$ are symmetries of $P\hat{Z}P$,
each eigenvalue of $P\hat{Z}P$ can be labelled by a reduced wavevector $\bkp{=}(k_x,k_y)$ in a reduced Brillouin zone $\textrm{rBZ}$, and eigenvalues that are continuously connected form a band. The corresponding eigenstates, known as Bloch-Wannier bands \cite{PhysRevB.89.115102,PhysRevB.83.035108}, are extended (in the $xy$-plane) as Bloch functions with crystal wavevector $\bm{k}_{\perp}$, but exponentially localized as a Wannier function in the $z$ direction. Such exponential localization makes the expectation value $\langle{\hat{Z}}\rangle,$ for a Bloch-Wannier eigenstate, well-defined, with $\langle{\hat{Z}}\rangle$ equalling an eigenvalue of $P\hat{Z}P$. Under translation in the $z$-direction by a lattice period (set to unity),  $P\hat{Z}P\rightarrow P(\hat{Z}+1)P$, hence each eigenvalue in a given $\bm{k}_{\perp}$ sector belongs to an  infinitely-extended Wannier-Stark ladder where adjacent rungs having unit spacing \cite{wannier_starkladder}. Generally, if the bulk valence energy band has rank $N$, the spectrum of $P\hat{Z}P$ (at each $\bm{k}_{\perp}$)  comprises $N$ Wannier-Stark ladders  -- each of them infinite in extent. For the Hopf insulator, $N=1$. 



Let us now consider a half-infinite slab geometry  with $\hat{Z}$ taking only positive values. We assume that a gap exists in the energy spectrum over the reduced Brillouin zone (rBZ) of the surface facet, and separates two orthogonal subspaces projected by $P$ (for occupied states) and $Q$ (for unoccupied states). (Such a gap is consistent with the possible existence of chiral hinge modes, because the surface geometry considered here is periodic without hinges.) Let $j=1,2,3\ldots$ label Bloch-Wannier bands with $j=1$ being closest to the surface, $j=2$ the next closest, and so on. Owing to the exponential decay of $P$ in the coordinate representation \cite{PhysRev.135.A685,PhysRevB.74.235111}, Bloch-Wannier bands sufficiently far from the surface are indistinguishable (up to exponentially small corrections) from Bloch-Wannier bands obtained in the above-described infinite geometry. We describe the `sufficiently far' Bloch-Wannier bands as \textit{bulk-like}, while Bloch-Wannier bands in the infinite geometry are \textit{bulk} Bloch-Wannier bands; every other Bloch-Wannier band is a \textit{surface} band. In the usual convention, we attach the additional qualifier `valence' and `conduction' to objects associated to the occupied and unoccupied subspaces respectively, e.g., valence bulk-like Bloch-Wannier bands. 

Let us denote the Chern number of the $j$-th Bloch-Wannier band in the occupied subspace as $C_{v}(+\hat{z},j)$. The \textit{faceted valence Chern number} $C_{v}(+\hat{z})$ is defined as the net Chern number of \textit{all} surface valence bands: $\sum_j C_{v}(+\hat{z},j)$, with the sum extending to the bulk-like region. This sum is uniquely defined because bulk-like Bloch-Wannier bands have zero Chern number, by assumption. Analogously, we can define $C_{c}(+\hat{z},j)$ and $C_{c}(+\hat{z})$ for the conduction band, $i.e.$, 
\begin{equation}
\label{eq: csv_csc}
\begin{array}{c}
C_{v}(+\hat{z})=\sum_{j=1}^{bulk}C_{v}(+\hat{z},j), 
\\
C_{c}(+\hat{z})=\sum_{j=1}^{bulk}C_{c}(+\hat{z},j).
\end{array}
\end{equation} 
In numerics, one can calculate the faceted valence/conduction Chern numbers by summing $j$ over half a slab, assuming the slab height is much greater than the exponential decay length of $P$ and $Q$. The definitions in \q{eq: csv_csc} are straightforwardly generalized to $C_{v,c}(\hat{\bm{n}})$ any insulating surface facet with normal vector $\hat{\bm{n}}$. Then, we define the \textit{faceted Chern number} of a surface as the sum of the faceted valence Chern number and the faceted conduction Chern number:
\begin{align}C(\hat{\bm{n}})=C_{v}(\hat{\bm{n}})+C_{c}(\hat{\bm{n}})=\chi, \label{eq:bulkboundary}
\end{align}
where the second equality relates the faceted Chern number to the bulk Hopf invariant $\chi$ [cf.\ Eq.\eqref{eq:hopfnumber}] -- a bulk-boundary correspondence proven in \ocite{alexandradinata2019actually}.

This correspondence suggests that the faceted valence Chern number $C_{v}(+\hat{z})$, just like the Hopf invariant, is only well-defined  if  the bulk valence energy band is unit-rank with trivial first Chern class. To appreciate this point, we offer an example with non-unit rank where $C_{v}$ is not well-defined. Our example is constructed by stacking decoupled, two-dimensional Chern insulators (in the $z$-direction) to form a three-dimensional superlattice with two layers per superlattice period, and with the Chern number alternating in sign between adjacent layers. The result is a topologically trivial insulator with two  Bloch-Wannier bands per lattice period in the z-direction, with the Chern number alternating in sign between adjacent Bloch-Wannier bands, as schematically shown in Fig. \ref{fig:schematic}(b). It follows that $C_v$ is not uniquely defined and depends on how the sum in \q{eq: csv_csc} is truncated: if  the number ($N$) of Bloch-Wannier bands included in the summation is even (see dotted square in Fig. \ref{fig:schematic} (b)), then $C_{v}=0$, while if $N$ is odd, $C_{v}=1.$

To close this subsection, let us illustrate the notions of  Bloch-Wannier bands, faceted Chern numbers, and the bulk-boundary correspondence for a tight-binding model of the Hopf insulator constructed by Moore, Ran, and Wen (MRW) \cite{PhysRevLett.101.186805}. The Hamiltonian of the MRW model is constructed from the well-known Hopf map:
\begin{equation}
\label{eq: MRW}
\begin{aligned}
&z=(z_1+iz_2,z_3+iz_4)^{T},
\\
&\mathbf{d}=z^{\dag}\bm{\tau}z, \ \bm{\tau}=(\tau_{x},\tau_{y},\tau_{z}),
\\
&H_{MRW}(\mathbf{k})=\mathbf{d}\cdot\bm{\tau},
\end{aligned}
\end{equation}
where $\tau_{x},\tau_{y},\tau_{z}$ are Pauli matrices, and
\begin{equation}
\label{eq:z1234}
\begin{aligned}
&z_{1}=\sin k_{x}, z_{2}=\sin k_{y}, z_{3}=\sin k_{z},
\\
&z_{4}=u-\cos k_{x}-\cos k_{y}-\cos k_{z}.
\end{aligned}
\end{equation}
If we take $1<u<3$, then the above model is a Hopf insulator with Hopf invariant $\chi=1$. In numerical calculations, we always use the MRW model with $u=2$ and $\chi=1$ unless otherwise specified. 

\begin{figure}[h]
\centering
\includegraphics[width=1\columnwidth]{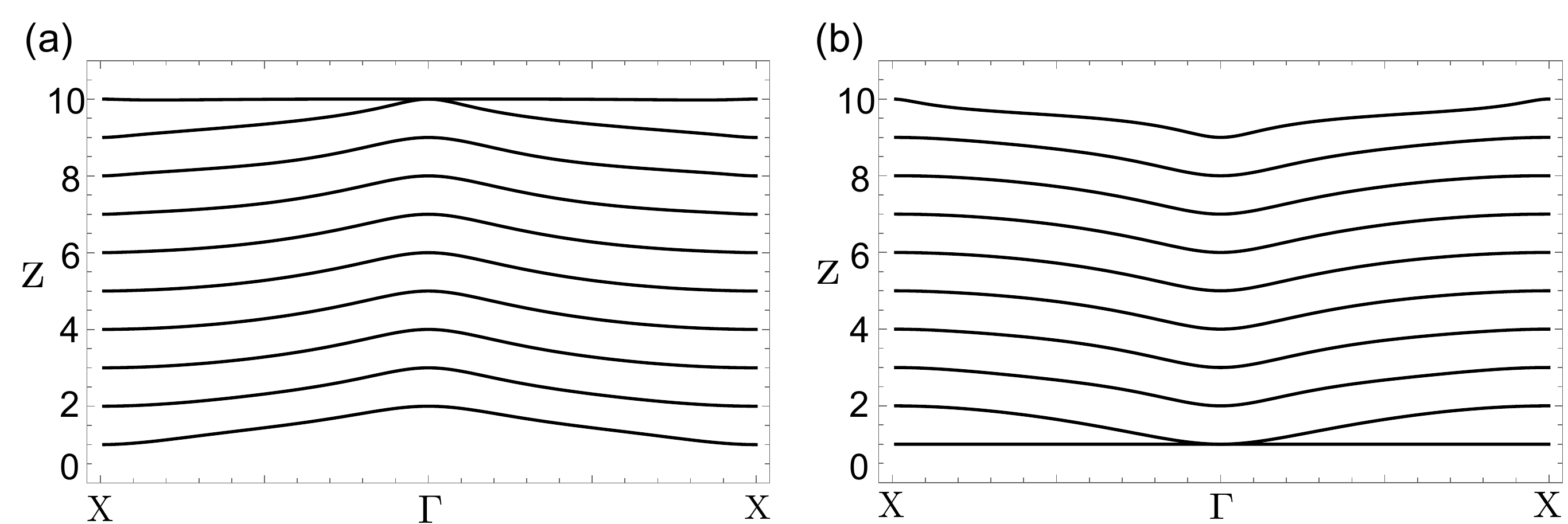}
\caption{Spectrum of (a) $P\hat{z}P$ and (b) $Q\hat{Z}Q$ for the MRW model with $u=2$ and $\chi=1$, for a finite slab with a width of ten unit cells in the $z$ direction. Except for Bloch-Wannier bands in the vicinity of the top and bottom facets, all other bands are bulk-like with near-unit separation.}
\label{fig:hbw_hopf}
\end{figure}

To calculate the Bloch-Wannier bands in the presence of a surface, we Fourier transform the Bloch Hamiltonian of the MRW model to obtain the position-space Hamiltonian. Then we terminate the hopping matrix elements across the surfaces of a finite slab that is open in  the $z$-direction and periodic in $x$ and $y$.  After modifying the surface Hamiltonian to gap out both the top and bottom surfaces, we diagonalize $P\hat{Z}P$ and $Q\hat{Z}Q$ to derive Bloch-Wannier bands as shown in Fig. \ref{fig:hbw_hopf}(a) and (b). If we sum the Chern number of the top five bands of $P\hat{z}P$ in Fig. \ref{fig:hbw_hopf}(a), we get $C_{v}(+\hat{z})=1$. If we sum the Chern number of the top five bands of $Q\hat{z}Q$ in Fig. \ref{fig:hbw_hopf}(b), we get $C_{c}(+\hat{z})=0$. Then, we have $C(+\hat{z})=C_{v}(+\hat{z})+C_{c}(+\hat{z})=1$, which is consistent with $\chi=1$, according to the bulk-boundary correspondence in Eq.\ \eqref{eq:bulkboundary}.  


\subsubsection{$C_{v}$ as a topological invariant}
\label{sec: csvtopo}

Let us prove that $C_{v}$ is  invariant under continuous deformations of the Hamiltonian that preserve the bulk translation symmetry as well as the energy gap separating valence and conduction subspaces (on a half-infinite geometry). 

The only way for $C_{v}$ to change is by the transference of Berry-curvature quanta to (or away from) the facet. If the energy gap is preserved, then there can be no transference between occupied and unoccupied states. What remains is to rule out transference between occupied states. Because Bloch-Wannier bands have differing spatial centers $\langle \hat{Z}\rangle$, the transference of a Berry-curvature quantum between neighboring Bloch-Wannier bands effectively moves the quantum in real space. Such transference can only occur if the spectral gap separating the two Bloch-Wannier bands closes then reopens; generically, the gap closes at a Dirac point. The transference of Berry quanta by Dirac-point touchings of Bloch-Wannier bands is sometimes referred to as an axion pump, and has been explored in Refs. \onlinecite{PhysRevLett.114.096401,PhysRevB.95.075137,PhysRevB.102.035166}  in the context of topological insulators whose bulk valence bands have rank larger than one. Yet, in our context, if bulk translation symmetry is preserved, it is not possible for bulk/bulk-like Bloch-Wannier bands (with unit rank) to touch, because they are always separated by the lattice period in the $z$-direction; by implication, the axion pump is forbidden.  An analogous argument proves that $C_{c}$ (for the unoccupied subspace) is also a topological invariant.

We can provide an alternative derivation of this result by assuming translation invariance in all three spatial directions. An adiabatic cycle for a bulk pump of Berry-curvature quanta corresponds to a nontrivial second Chern number  for the Hamiltonian $H(\bk,\lambda)$\cite{PhysRevLett.114.096401,alexandradinata2019actually}, where $\lambda$ is a cyclic parameter for a family of Bloch Hamiltonians $H(\bk)$ defined over the 3D Brillouin zone. The second Chern number is defined as a four-dimensional integral of the second Chern character. Generally, the $n$'th Chern character is given by
\e{ ch_n(F) = \tr \left[ \f{1}{n!} \bigg(\f{iF}{2\pi}\bigg)^n\right],}
with $F^2=F\wedge F$ the exterior product of the  Berry field strength. However, the second Chern number is always trivial for a non-degenerate band 
spanned by a single Bloch function [of $(\bk,\lambda)$] that is continuous with respect to $\bk$. This follows from a relation between the first two Chern characters and the second Chern class $c_2$:\cite{stone_mathematicsforphysics}
\e{ c_2(F) = \half ch_1(F)\wedge ch_1(F) -ch_2(F), \la{characterclass}}
and the triviality of the second Chern class (i.e., $\int d^3k d\lambda \, c_2{=}0$) for line bundles \cite{hatcher_vectorbundlesKtheory}. Note $\int d^3k d\lambda\, ch_1{\wedge} ch_1{=}0$ because,  by assumption, the first Chern numbers vanish in the $k_xk_y$, $k_yk_z$ and $k_zk_x$ planes. This illustrates that adiabatic deformations of the Bloch Hamiltonian will not transfer any Berry curvature quanta away from the facet. A similar argument was used in \ocite{alexandradinata2019actually} to demonstrate the topological invariance of the faceted Chern number $C$ [cf.\ \q{eq: csv_csc}].

\subsection{Surface anomalous Hall conductivity}
\label{sec:sahc_hopf}


The half-integer quantization of the faceted compressibility of the Hopf insulator [cf.\ \q{eq: dmdmu_hopf}] can be proven once we relate the geometric SAHC to the bulk topological invariant $\chi$ [cf. \ \q{eq:hopfnumber}] and the faceted valence Chern number $C_{v}$ [cf. \ \q{eq: csv_csc}] through \q{geometricsahchopf}. As  preliminary steps, we will first review some salient aspects of the general theory of surface Hall conductivity in \s{sec:sahcgeneral}, then  generalize Laughlin's argument to surface facets in \s{sec:sahctrivial}. This general theory is subsequently applied to the Hopf insulator  in \s{sec:sahchopf}.  

\subsubsection{Review of the general theory of magnetoelectric polarizability and SAHC}\la{sec:sahcgeneral}


For an insulator with a gapped surface facet (with normal vector $\hat{\bm{n}}$), the application of an electric field results in a surface Hall current: $\bm{j}_{\hat{\bm{n}}}=\sigma_{\hat{\bm{n}}}\hat{\bm{n}}\times \bm{E}$. The total SAHC $\sigma_{\hat{\bm{n}}}$ can be separated into geometric ($\sigma_{\hat{\bm{n}}}^{g}$) and non-geometric contributions\cite{qi2008topological,essin2009magnetoelectric,essin2010orbital,malashevich2010theory,thonhauser2011theory,sitte2012topological,rauch2018geometric,vanderbilt_2018,PhysRevB.98.245117,10.21468/SciPostPhys.6.4.046}:
\e{
\label{eq: surfhall_general}
\sigma_{\hat{\bm{n}}}\eq\sigma_{\hat{\bm{n}}}^{g}-\alpha_{iso}^{cg}+\frac{1}{2}\widetilde{\alpha}_{ab}\hat{n}_{a}\hat{n}_{b},\lin
\sigma_{\hat{\bm{n}}}^{g}\eq \left(-\frac{\theta}{2\pi}+m_{s}(\hat{\bm{n}}) \right)\frac{e^{2}}{h}, \as m_{s}\in\Z.
} The geometric contribution comprises two terms, the first of which is proportional to the axion angle $\theta.$ The value of $\theta$ is expressible as a Brillouin-zone integral of the Chern-Simons three-form of the Berry gauge field \cite{qi2008topological}. For a unit-rank valence energy band, the Chern-Simons three-form is Abelian, and the axion angle reduces to $\pi$ multiplied by the right-hand side of \q{eq:hopfnumber}, i.e., it is $\pi \chi.$ The second term in  $\sigma_{\hat{\bm{n}}}^{g}$ is an integer multiple of $e^2/h$ that depends on the surface preparation. The non-geometric contribution to $\sigma_{\hat{\bm{n}}}$ also comprises two terms, one being facet-independent (the ``cross-gap" term
$\alpha_{iso}^{cg}$), and the other facet-dependent. Analytic expressions for both these terms can be found in Refs. \onlinecite{essin2010orbital,malashevich2010theory}, but will not be the focus of this work.


\subsubsection{SAHC of insulators with trivial bulk Chern invariants: a generalized Laughlin's argument}\la{sec:sahctrivial}

The differential SAHC between any two adjacent facets is an integer multiple of $e^2/h$, with this integer equal to the net chirality of hinge modes localized to the interface of the two facets. This follows from a generalization of Laughlin's argument to surface facets that we now develop. 

\begin{figure}[h]
\centering
\includegraphics[width=1\columnwidth]{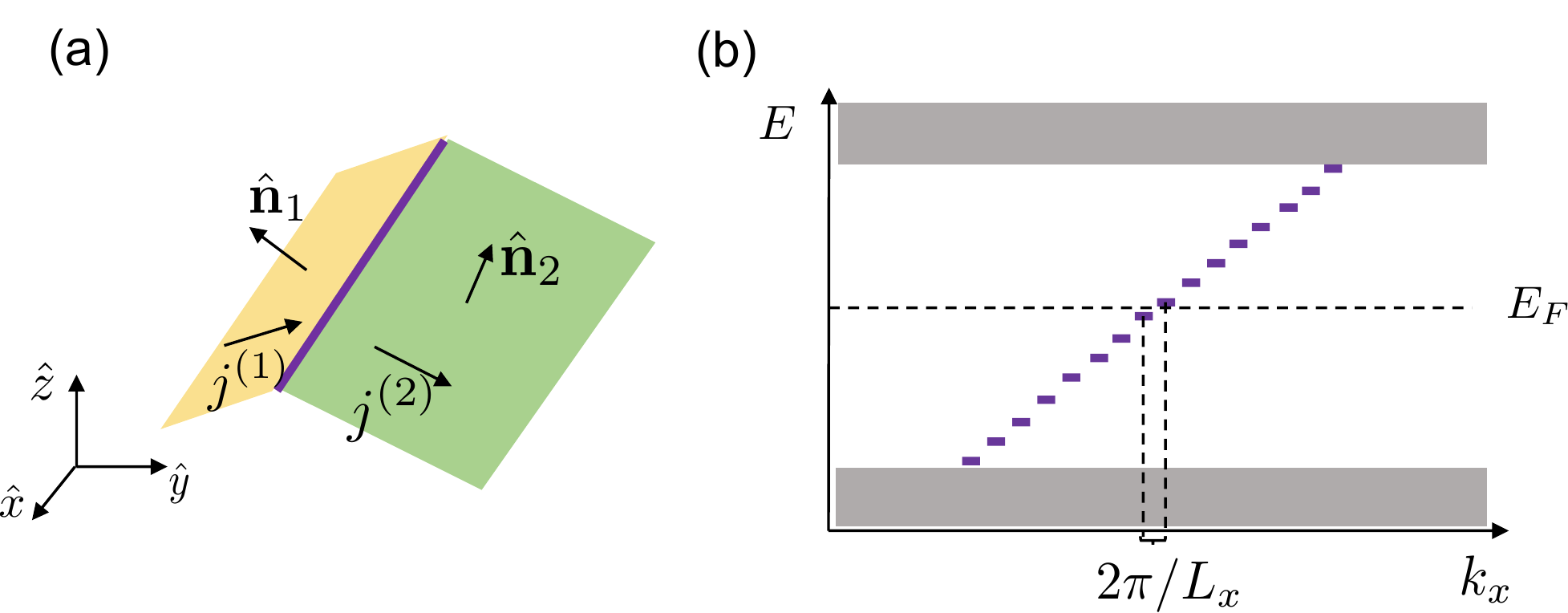}
\caption{(a) Two adjacent facets with normal vectors $\hat{\bm{n}}_{1}$ and $\hat{\bm{n}}_{2}$. Their interface is emphasized by a purple colored line. (b) Illustration of spectral flow for a single chiral hinge mode. The purple dashes indicate hinge states with wavenumber $k_x$ that takes discrete values owing to finite-size quantization.}
\label{fig:Laughlin}
\end{figure}

Laughlin's original argument was applied to prove the integer quantization of the Hall conductivity for two-dimensional electron gases \cite{PhysRevB.23.5632}, using flux threaded in a cylindrical geometry. Our generalized argument is based on the geometry illustrated in Fig.  \ref{fig:Laughlin}(a): two adjacent surface facets with normal vectors $\hat{\bm{n}}_{1}$ and $\hat{\bm{n}}_{2}$, having an interfacial ``hinge" of length $L_x$ parallel to $+\hat{x}$. Periodic boundary conditions are imposed in the $x$ direction, while each facet is assumed to be semi-infinite in the direction orthogonal to $x$. 


Let us introduce a time-dependent flux $\Phi(t)$ into the Hamiltonian by the minimal coupling $p_x \ri p_x - (e/c)\Phi/L$, with $p_x$ the canonical momentum operator. One can remove the flux from the Hamiltonian via a unitary transformation which twists the boundary conditions of single-body wave functions as: $\psi(x,y,z)=\exp\left(-i \Phi/\Phi_0\right)\psi(x+L_x,y,z)$, with $\Phi_0=hc/e$ the flux quantum. The spectrum of the Hamiltonian is identical for any $\Phi$ that is an integer multiple of $\Phi_0$. However, it is possible that a filled Fermi sea ground state (at $\Phi=0$) is adiabatically connected to an excited many-body state after a flux quantum is threaded (at $\Phi=\Phi_0$). This happens only if there are states (near the Fermi level) that are extended in the $x$-direction; such extension makes them sensitive to twisting of the boundary condition along $x$. Since we have assumed the bulk and surfaces are gapped, any extended state (near the Fermi level) will be a quasi-1D Bloch wave localized to the hinge. If hinge modes crossing the Fermi level have a  net chirality for propagation in the $x$-direction, then the many-body state at $\Phi=\Phi_0$ is excited when compared to $\Phi=0$.   For illustration, if the spectrum contains a single chiral hinge mode (illustrated in Fig. \ref{fig:Laughlin} (b)), then all single-body states are advanced in quasimomentum by $k_x\ri k_x+2\pi/L_{x}$, resulting in one additional filled state just above the Fermi level.

More generally, let the net chirality of the hinge modes be $\Delta N=N_R-N_L$, implying that the number of occupied hinge states increases by $\Delta N$ upon inserting of a flux quantum. By particle conservation, these states must originate from a net flow of current transverse to the hinge: $(-e)\Delta N=L\int_0^{T_0}(j^{(1)}-j^{(2)})dt$, with $j^{(i)}$ the areal current density on the $i$-th facet. We note that there is no current contributed from the bulk, because the bulk is insulating, and the bulk Chern invariants are assumed to vanish.  A time-dependent flux $\Phi(t)=LA_x(t)$ can be viewed as proportional to a time-dependent electromagnetic vector potential, which implies a non-vanishing electric field: $E_x=(-1/c)\partial A_x/\partial t$. Assuming $E_x$ is constant in time, then  $\Phi(T_0)=\Phi_0$ at time $T_0=- h/eE_xL$. Substituting this time scale into the above expression for $\Delta N$, and further identifying $j^{(i)}=\sigma_{\hat{\bm{n}}_{i}}E_x$ as a Hall current (since both surfaces are gapped), we derive that the difference in Hall conductivity is determined by the net chirality: $\sigma_{\hat{\bm{n}}_{1}}-\sigma_{\hat{\bm{n}}_{2}}=(e^2/h) \Delta N$. This completes our argument.

We can now use this result to interpret the general expression for the SAHC in Eq. \ref{eq: surfhall_general}. First, we argue that  in three-dimensional insulators whose bulk valence band has trivial Chern class, the anisotropic term $\frac{1}{2} \widetilde{\alpha}_{a b} \hat{n}_{a} \hat{n}_{b}$ in Eq.  \eqref{eq: surfhall_general} must vanish. This follows because on one hand, (i) if $\widetilde{\alpha}$ were nonzero, $\frac{1}{2} \widetilde{\alpha}_{a b} \hat{n}_{a} \hat{n}_{b}$ can in principle take arbitrary real values depending on the facet orientation $\hat{\bm{n}}$, implying that the differential $\frac{1}{2} \widetilde{\alpha}_{a b} \hat{n}_{a} \hat{n}_{b}$ between two facets can take arbitrary real values. On the other hand, (ii) we just proved that the differential SAHC between two adjacent facets must be an integer multiple of $e^2/h$. Since the real-valued arbitrariness of (i) is incompatible with the integer-valued quantization of (ii), our previous assumption of $\widetilde{\alpha}\neq 0$ cannot be true.

Secondly, we note that $\Delta N$ can be identified with the differential $m_s$ between two adjacent facets. After all, having just argued that $\widetilde{\alpha}=0$, $m_s$ is the only remaining nonzero quantity in \q{eq: surfhall_general} that depends on the facet orientation.

\subsubsection{SAHC of the Hopf insulator}\la{sec:sahchopf}

 Specializing the general expression of the SAHC in \q{eq: surfhall_general} to the case of the Hopf insulator, we obtain
\begin{equation}
\label{eq: surfhall}
\begin{array}{c}
\sigma_{\hat{\bm{n}}}=\sigma_{\hat{\bm{n}}}^g-\alpha_{iso}^{cg}, \as \sigma_{\hat{\bm{n}}}^g=\left(-\frac{\chi}{2}+m_{s}(\hat{\bm{n}})\right)\frac{e^2}{h}. 
\end{array}
\end{equation} We have replaced the (Abelian) axion theta angle by the Hopf invariant: $\theta_{Abelian}=\pi\chi$, in accordance with Eq.
\eqref{eq:hopfnumber}. (Note that $\chi$ is gauge-invariant, unlike the non-Abelian version where $\theta_{NA}\equiv \theta_{NA} +2\pi$.) Moreover,  the term involving $\tilde{\alpha}$ vanishes as an implication of the generalized Laughlin argument given above.

Next we will argue that we can identify the surface-dependent integer invariant $m_s(\hat{\bm{n}})$ with the faceted valence Chern number (c.f.  \q{eq: csv_csc}): 
\e{m_s(\hat{\bm{n}})=C_{v}(\hat{\bm{n}}).} 
To prove this identity, consider a half-infinite slab of the Hopf insulator with surface-normal vector $\hat{\bm{n}},$ and total SAHC  $\sigma_{\hat{\bm{n}}}$. 
The surface valence bands have the net Chern number $C_{v}(\hat{\bm{n}}),$ which depends on details of the surface termination. As a thought experiment, let us absorb a two-dimensional Chern-insulating layer onto the surface which has the opposite Chern number $-C_{v}(\hat{\bm{n}})$ in its valence subspace. We do not, however, introduce any Hamiltonian matrix elements between the absorbed layer and the original system. The combined system is then a half-infinite, Hopf-insulating slab whose occupied subspace can be spanned by a set of exponentially-localized Wannier functions. Then by following the derivations in Refs. \onlinecite{malashevich2010theory,thonhauser2011theory} (which apply only to systems with a Wannier representation), we would deduce that the SAHC of the combined system equals $-\chi/2 (e^2/h)-\alpha_{iso}^{cg}$. In the assumed absence of coupling between the Chern-insulating layer and the original slab, the SAHC of the combined system must equal the sum of the SAHC of the constituent systems, thus
\begin{equation}
\sigma_{\hat{\bm{n}}}-C_{v}(\hat{\mathbf{n}})\frac{e^2}{h}=-\frac{\chi}{2} \frac{e^2}{h}-\alpha_{iso}^{cg}.
\end{equation}
Comparing this to the first equation in Eq.\eqref{eq: surfhall}, we deduce $m_{s}(\hat{\bm{n}})=C_{v}(\hat{\bm{n}})$, and hence, for the Hopf insulator 
\begin{equation}
 \sigma_{\hat{\bm{n}}}^g=\left(-\frac{\chi}{2}+C_{v}(\hat{\bm{n}})\right)\frac{e^2}{h}. \end{equation}

We remark that for Hopf insulators whose  bulk Hamiltonian $H(\bk)$ is traceless, as exemplified by  the MRW model in Eq. \eqref{eq: MRW}, the SAHC is purely geometric. This is because the traceless condition ensures that  the `{degeneracy}' and `{reflection}' conditions [discussed in Ref \onlinecite{essin2010orbital}] are satisfied, which imply the vanishing of the non-geometric, `cross-gap'  SAHC. 

\subsection{Chiral hinge modes and their symmetry protection }
\label{sec: chiral_difference}

\begin{figure}[h]
\centering
\includegraphics[width=1\columnwidth]{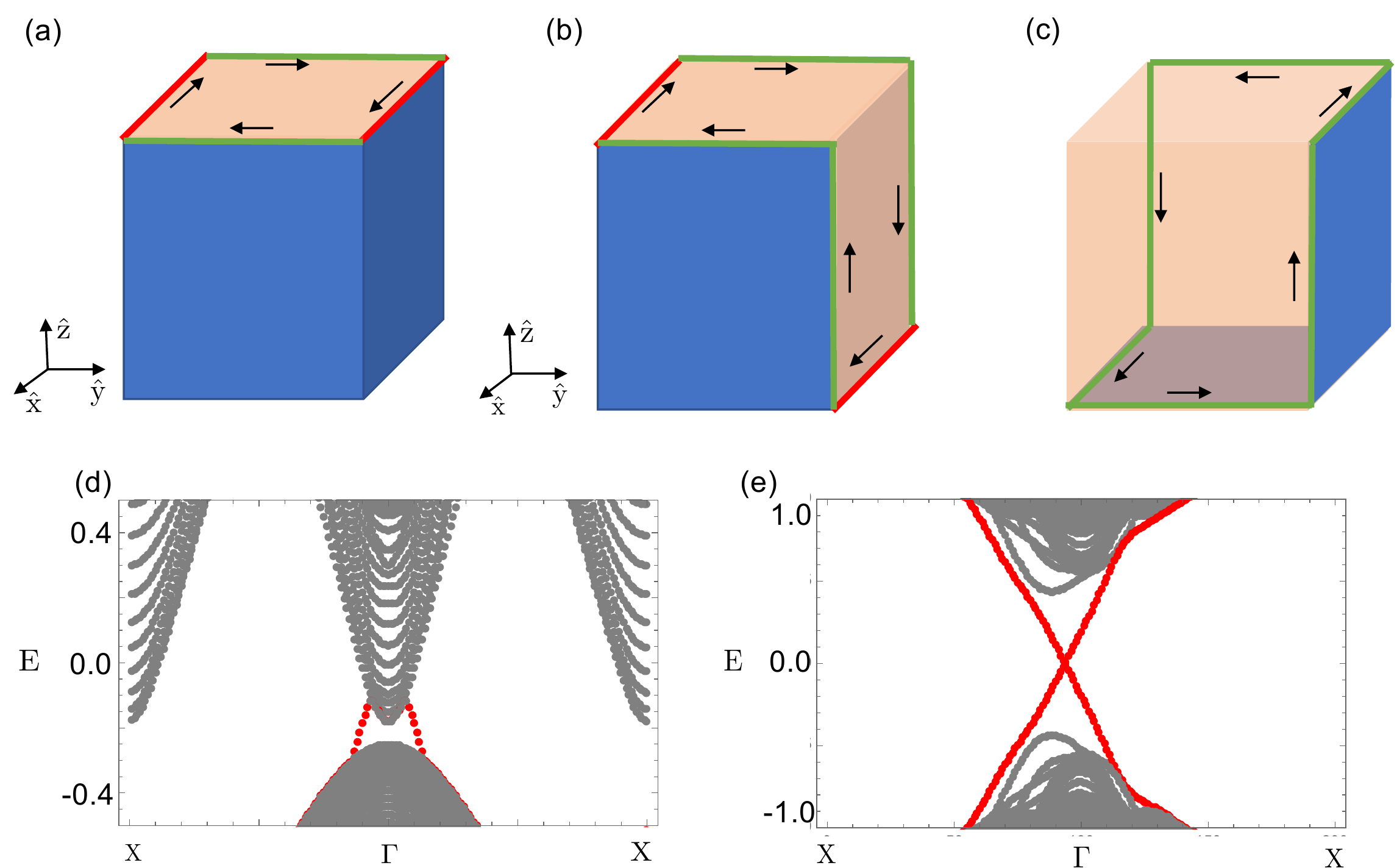}
\caption{ (a-c) shows the possible networks of hinge modes for a cubic Hopf-insulating crystalline with bulk Hopf invariant $\chi=1$. Facets with $C_v=1$ ($C_v=0$) are colored light orange (blue, respectively). Black arrows represent the the group velocity of chiral hinge modes at the Fermi level. The particle-hole inversion symmetry of \q{phisymmetry} is imposed in panels (b) and (c), but not in (a). The dispersion of hinge modes localized to the red-highlighted hinges in panel (a) [resp.\ (b)] are plotted in panel (d) [resp.\ (e)]. Green-colored hinges also hosts hinge modes which are, however, not shown in an energy-momentum plot.
Simulations for (d-e) were performed for the MRW model ($u=2$) on a finite slab that is periodic in the $x$ direction, with a width of 10 unit cells in the $z$-direction, and a width of 40 unit cells in the $y$-direction.}
\label{fig:hinge_hopf}
\end{figure}

We have just determined that the net chirality of hinge modes separating two surface facets equals the differential $C_{v}$ between the two adjacent facets. Given a Hopf-insulating crystallite with a fixed bulk invariant $\chi,$  it is possible to manufacture a surface potential that leads to many choices of $\{C_{v}(\hat{\bn})\}_{\hat{\bn}}$, and by implication many possible networks of hinge modes. Three representative examples consistent with half filling of the MRW model are illustrated in \fig{fig:hinge_hopf}(a-c), with blue indicating a facet with $C_{v}=+1$, light orange indicating a facet with $C_{v}=0$, and  red indicating a unidirectional hinge mode. These three simulations differ only in the surface preparations, while their bulk is identically given by the MRW model [cf.\ \q{eq: MRW}] with parameter $u=2,$ and bulk invariant $\chi=1$. For the two examples in \fig{fig:hinge_hopf}(a-b), the dispersion of the  hinge modes propagating in the $\pm x$ direction are plotted in   \fig{fig:hinge_hopf}(d-e) respectively.

\begin{figure*}[t]
\centering
\includegraphics[width=2\columnwidth]{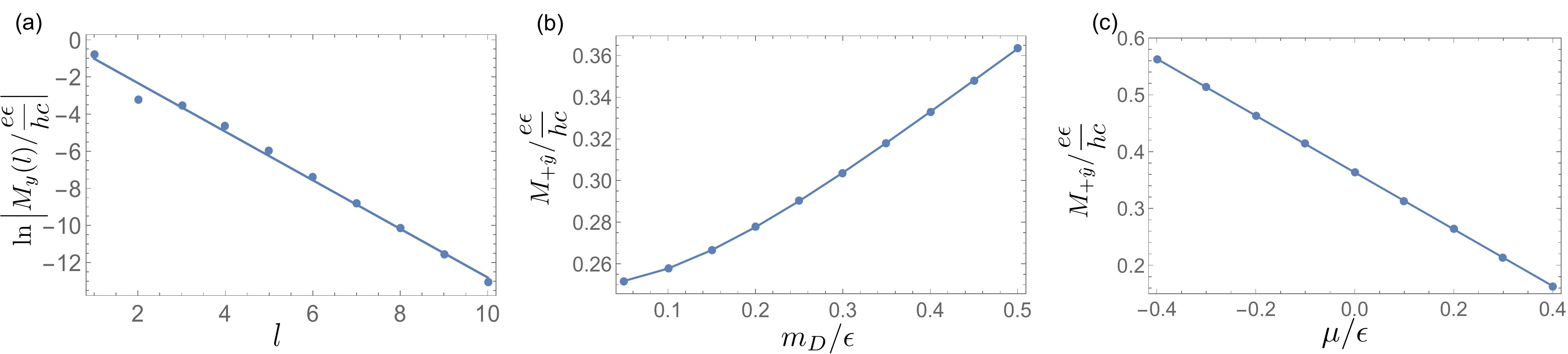}
\caption{ Simulations on a finite slab that is open in the $y$-direction, with 20 unit cells. Figure (a) plots $\ln\left|M_{y}(l)/(e\epsilon/h c)\right|$ \emph{v.s.} the  layer index $l$, with $l=1$ closest to the surface termination; we have chosen the surface Dirac mass to be $m_{D}=0.1\epsilon$. Exponential decay of the magnetization away from the surface is observed.  (b) plots $M_{+\hat{y}}$ (at $\mu=0$) as a function of $m_{D}$. (c) plots $M_{+\hat{y}}$ \emph{vs} $\mu$ with $m_{D}=0.5\epsilon$. The constant slope as a function of $\mu$ indicates a quantized compressibility.}
\label{fig:mag}
\end{figure*}


Interestingly, it is possible that all facets have identical $C_{v}$ such that there are no chiral hinge modes at all. Hence, one might consider if a symmetry exists that might prevents this possibility and thus guarantee higher-order topology with robust hinge modes. As a candidate for such a symmetry, let us consider the following transformation of  annihilation to creation operators 
\e{\mathcal{C}^{\prime}: c_{\bm{R},a}\ri \epsilon_{ba}c_{-\bm{R},b}^{\dag}.\la{phisymmetry}}
Here, $c_{\bm{R},a}^{\dag}$ creates an electron of orbital character $a\in \{1,2\}$ at lattice site $\bm{R}$, and $\epsilon_{ab}$ is the two-dimensional Levi-Civita tensor.
Since $\mathcal{C}^{\prime}$ is the composition of a spatial inversion $\cali: c_{\bm{R},a}\ri c_{-\bm{R},a}$ and a particle-hole conjugation $\calp: c_{\bm{R},a}\ri \epsilon_{ba}c_{\bm{R},b}^{\dag}$ (with pseudo-spin flip), we refer to $\mathcal{C}^{\prime}=\calp \cali$ as a \textit{particle-hole inversion} symmetry. Note that neither $\calp$ nor $\cali$ symmetry is individually imposed in the following discussion. The $\mathcal{C}^{\prime}$ symmetry of the second-quantized Hamiltonian translates to the following antisymmetry of the first-quantized Hamiltonian in the momentum representation:
\e{\tau_{y} H^{\star}(\bk)\tau_{y} = -H(\bk), \as \bk \in BZ\label{eq:antisymmetry},}
where $H^{\star}(\bk)$ means the complex conjugate of the matrix $H(\bk)$.
This constraint is satisfied by any traceless, two-by-two matrix Hamiltonian. 

We propose that $\mathcal{C}^{\prime}$ symmetry guarantees hinge modes for Hopf insulators with odd bulk invariant $\chi$, by ensuring that $C_{v}$ on inversion-related facets are distinct, as  illustrated in \fig{fig:hinge_hopf}(b) and (e). To derive the necessity of hinge modes, let us express $C=C_{v}+C_{c}$ as an integral (over the rBZ) of the Abelian Berry curvature of the surface valence (and conduction) bands:
\begin{equation}
\label{eq: surfchern2}
\begin{aligned}
C(\hat{\bm{n}})&=\frac{1}{2\pi}\int_{rBZ}d^{2}k \big\{Tr\left[\bm{\mathcal{F}}(\bm{k})_{\hat{\bm{n}},v}\cdot\hat{\bm{n}}\right]
\\
&+Tr\left[\bm{\mathcal{F}}(\bm{k})_{\hat{\bm{n}},c}\cdot\hat{\bm{n}}\right]\big\}= C_{v}(\hat{\bm{n}})+C_{c}(\hat{\bm{n}}).
\end{aligned}
\end{equation}
We can understand how $\mathcal{C}^{\prime}$ constrains the Berry curvature by decomposing its action as described above. Since the Berry curvature is a pseudo-vector, the spatial inversion transformation sends $\bm{\mathcal{F}}(\bm{k})_{\hat{\bm{n}},v}$ to $\bm{\mathcal{F}}(-\bm{k})_{-\hat{\bm{n}},v}$. Next, the particle-hole conjugation transformation
subsequently sends $\bm{\mathcal{F}}(-\bm{k})_{-\hat{\bm{n}},v}$ to $-\bm{\mathcal{F}}(\bm{k})_{-\hat{\bm{n}},c}$. The minus sign is due to the anti-unitary action of particle-hole conjugation in $\bk$-space. Thus, $\mathcal{C}^{\prime}$ ultimately constrains $\bm{\mathcal{F}}(\bm{k})_{\hat{\bm{n}},v}= -\bm{\mathcal{F}}(\bm{k})_{-\hat{\bm{n}},c}$, and therefore
\begin{equation}
\label{eq:SCconstraint}
\begin{aligned}
&C_{v}(\hat{\bm{n}})=\frac{1}{2\pi}\int_{rBZ}d^{2}k Tr\left[\bm{\mathcal{F}}(\bm{k})_{\hat{\bm{n}},v}\cdot\hat{\bm{n}}\right]
\\
&= \frac{1}{2\pi}\int_{rBZ}d^{2}k Tr\left[\bm{\mathcal{F}}(\bm{k})_{-\hat{\bm{n}},c}\cdot\left(-\hat{\bm{n}}\right)\right]=C_{c}(-\hat{\bm{n}}).
\end{aligned}
\end{equation}
Combining this with the bulk-boundary correspondence [cf.\ \q{eq:bulkboundary}], we find that if $\chi$ is odd, the parities of $C_{v}(\hat{\bm{n}})$ and $C_{v}(-\hat{\bm{n}})$ are non-identical. This would indicate that some change of the SAHC is required over neighboring facets, and hence there will be localized hinge modes. However, for $\chi$ even,  it is possible that $C_{v}(\hat{\bm{n}})= C_{v}(-\hat{\bm{n}})=\chi/2$ for all $\hat{\bm{n}}$, implying no generic, robust hinge modes.

Upon  enlargement of  the Hilbert space by addition of trivial bands, it was discovered that Hopf insulators (with odd Hopf invariants) remain homotopically inequivalent to a trivial insulator --  assuming an antisymmetry constraint that is a higher-rank generalization of \q{eq:antisymmetry} \cite{Liu2017symmetry}. Whether or not these higher-rank insulators are also higher-order topological insulators is a question we leave for future investigation.

\subsection{Numerical simulation of the magnetic compressibility}\la{sec:numericcompress}

For the MRW model, we will numerically  demonstrate that the layer-resolved magnetization is indeed exponentially-localized to a surface facet, and sensitive to details of the surface potential. On the other hand, the faceted magnetic compressibility $dM_{\hat{\bm{n}}}/d\mu$ is a half-integer-valued topological invariant if the  Hopf invariant $\chi$ is odd, no matter the details of the surface termination.



Beginning with the position-space representation of the MRW model with parameters ($u=2,\chi=1$), we diagonalize the Hamiltonian with a surface termination  in the $+\hat{y}$ direction, by abruptly truncating all Hamiltonian matrix elements that extend across the surface.  With this prescription, the surface energy bands touch at one Dirac point owing to an artifact of the simplified model. To gap the surface, we split this Dirac-point degeneracy by adding the Hamiltonian term: $m_{D}(\mathcal{P}_{l=2}-\mathcal{P}_{l=1})$, where $m_D$ can be interpreted as a Dirac mass, and $\mathcal{P}_{l=1,2}$ projects to the first and second layers (counting from the surface termination). We then fix the Fermi level in the resultant energy gap, and compute $M_{y}(l)$ on the surface with normal vector $+\hat{y}$ through Eq.\eqref{eq: Mghf2} (with $z$ replaced by $y$). Fig. \ref{fig:mag}(a) plots the absolute value $|M_{y}(l)|$ vs $l$, for the Dirac mass $m_{D}= 0.1 \epsilon$, where $\epsilon$ is half the bulk energy gap [see  Fig. \ref{fig:schematic}(a)]. This simulation demonstrates that the faceted orbital magnetization is indeed exponentially localized at the surface. What is more, as shown in Fig. \ref{fig:mag}(b), the faceted orbital magnetization on the surface with normal vector $+\hat{y}$ can be tuned continuously by varying the Dirac mass $m_{D}$. Since the Dirac mass is  experimentally tunable (at least  on the surface of a $\Z_2$ topological insulator \cite{chen2010massive}), we see that, in principle, the faceted orbital magnetization is tunable by surface manipulations. 

In contrast, Fig. \ref{fig:mag}(c) shows that $M_{+\hat{y}}$ depends linearly on $\mu$ with slope equal to  $\sigma_{+\hat{y}}^{g}=(1/2) (e^{2}/h)$. This is consistent with  Eq. \eqref{eq: surfhall} which states that the slope equals $(C_v-\chi/2) e^2/h$; here, $\chi=1$ follows from our choice of the bulk Hamiltonian parameter $u=2$, and  $C_{v}=1$ on the surface facet with normal vector $+\hat{y}$, due to the specific nature of our surface termination. 


\section{Topomagnetic insulators (and beyond)}\la{sec:topomagnet}

The half-integer-quantized faceted compressibility is not a unique property of the Hopf insulator; rather, it applies to any insulator whose geometric SAHC is a half-integer multiple of $e^2/h$, according to the general relation between compressibility and surface anomalous Hall conductivity  in \q{eq: dmdmu}. This class of insulators, sometimes referred to as axion insulators, are exemplified by (i)  the three-dimensional $\Z_2$ topological insulators \cite{PhysRevLett.98.106803,qi2008topological} with gapped surface states \cite{qi2008topological,sitte2012topological}, (ii) the 3D, second-order topological insulators protected by a composition  of time reversal with four-fold rotation ($TC_4$) \cite{schindler2018higher}, and (iii) certain anti-ferromagnetic topological insulators with bulk axion angle $\theta=\pi$ and with gapped surfaces \cite{PhysRevB.81.245209,PhysRevLett.122.256402}. 

In examples (i-ii), the bulk magnetization vanishes owing to symmetry -- time reversal in the first case, and $TC_4$ in the second. This is because both spin and orbital magnetization [cf.\ \s{sec: localM}] transform as a pseudo-vector under crystallographic spatial transformations, and are moreover odd under time reversal, as shown in Appendix \ref{app: Mtransform}. In example (iii),  the bulk magnetization vanishes because the magnetic order is anti-ferromagnetic. (i-iii) represent a counter-intuitive phase of matter which is insulating  (in the bulk and on surface facets) with vanishing bulk magnetization under zero field, but has a half-integer-quantized magnetic compressibility on its surface -- we call this a topomagnetic insulator. Having zero bulk magnetization renders the faceted magnetization more readily measurable, with techniques that are tentatively proposed in  \s{sec: conclusions}.


\subsection{The Hopf insulator as an anisotropic analog of the topomagnetic insulator}
\label{sec: mag}

Not all axion insulators are topomagnetic, e.g., time-reversal-broken axion insulators protected by spatial inversion symmetry \cite{PhysRevB.83.245132,qi2008topological,PhysRevB.98.245117,essin2009magnetoelectric,PhysRevB.86.115112} generally have nonvanishing bulk magnetization. The Hopf insulator is another axion insulator that is not topomagnetic,  however certain  symmetries render its bulk magnetization to vanish in two of three independent spatial directions. In this sense, the Hopf insulator realizes an anisotropic analog of a topomagnetic insulator.

The simplest symmetry that would constrain the bulk magnetization to vanish is time reversal. However, this symmetry is incompatible with the Hopf invariant $\chi \neq 0$. This is because $\chi$, being an integral of the Chern-Simons three-form [cf.\ \q{eq:hopfnumber}], transforms like a pseudoscalar under crystallographic spatial transformations, and is odd under time reversal.

The best alternative one can achieve for the Hopf insulator is to impose symmetry such that the bulk magnetization vanishes in two independent directions. For example, time reversal $\hat{T}$ composed with the reflection  $\hat{m}_{x}: x\ri -x,$  constrains the $x$-component of the magnetization to vanish, and is also compatible with $\chi \neq 0$. It is possible to simultaneously  have $\hat{T}\hat{m}_{x}$ and $\hat{T}\hat{m}_{y}$ symmetries, hence forbidding magnetization in the $x$ and $y$-directions.
However, if $\hat{T}\hat{m}_{z}$ is further imposed, the product of these three operations inverts $(x,y,z,t) \ri (-x,-y,-z,-t)$, which ensures the Berry curvature vanishes at each $\bk$, hence $\chi=0$. Incidentally, the MRW model [cf.\ Eq. \eqref{eq: MRW}] is manifestly symmetric under $\hat{T}\hat{m}_x$ and $\hat{T}\hat{m}_y$: 
\begin{equation}
\label{eq: tmxtmy}
\begin{array}{c}
\tau_{z} H_{MRW}^{\star}(\bm{k})\tau_{z}=H_{MRW}(k_{x},-k_{y},-k_{z}),
\\
H_{MRW}^{\star}(\bm{k})=H_{MRW}(-k_{x},k_{y},-k_{z}),
\end{array}
\end{equation}
making the MRW phase an anisotropic topomagnetic insulator. 

As a curious sidenote, because the MRW model Hamiltonian is traceless, it possesses an antisymmetry  [cf. Eq. \eqref{eq:antisymmetry}] that ensures that the orbital (but not the spin) magnetization vanishes in all directions. This follows from a more  general result: any three-dimensional insulator with trivial bulk Chern invariants and the following antisymmetry \e{UH^{\star}(\mathbf{k})U^{\dag}=-H(\mathbf{k}), \as \dg{U}=U^{-1}, \la{generalantisymmetry}}
must have vanishing extensive orbital magnetization in the bulk. This statement holds for Hamiltonians of any rank, and in particular it applies to  Pauli-matrix Hamiltonians, as proven in Appendix \ref{app: proof_hb}.

\section{Summary and outlook}
\label{sec: conclusions}


We have investigated the  subextensive magnetic moment $m$ contributed by crystalline facets of a three-dimensional insulator in the trivial first Chern class, assuming the existence of a bulk and surface energy gap. This magnetic moment has both dynamical and geometric contributions, where the latter is a product of the chemical potential $\mu$ with a momentum-space integral of the Berry curvature (for states localized to the facet). Consequently, the magnetic compressibility $A^{-1}dm/d\mu$ (contributed by one facet, per unit facet area $A$) is $e/hc$ times the geometric part of the surface anomalous Hall conductivity. In some cases the geometric contribution is restricted to a half-integer multiple of $e/hc$, owing either to symmetry or to Hilbert-space restrictions.

Geometry is ever linked with topology: two adjacent facets with differing geometrical conductivity must host chiral hinge mode(s) at their interface, where the net chirality is an integer-valued topological invariant. Thus we establish a connection between quantized surface magnetism and higher-order topology. When half-integer-quantization of $A^{-1}dm/d\mu$ is combined with a vanishing bulk magnetization, such a phase of matter is referred to as a topomagnetic insulator. 

An advantage of topomagnetic insulator is that the faceted magnetization is more readily measurable if the bulk contribution vanishes. Since the magnetization contributed by a single facet has never been measured (to our knowledge), we speculate on three existing experimental techniques that might be augmented to do the job:

\noi{i} Torque magnetometry couples to the total moment of a crystallite \cite{shoenberg}, which is a sum of contributions from the bulk and from all facets. The former contribution vanishes by assumption. Note that the contribution from opposite facets of a crystallite generically do not cancel out, as exemplified by our Hopf case study.

\noi{ii} Magnetic neutron diffraction  probes the magnetization \cite{hirst_magnetization} of a single facet,  assuming the cross-sectional area of the neutron beam is smaller than the faceted area. While the penetration depth of a neutron beam is typically much larger than the penetration depth of surface states, there is no need to isolate the faceted contribution if the bulk magnetization vanishes. 

\noi{iii} Measuring second harmonic generation is also a surface-sensitive diagnostic of magnetization in centrosymmetric materials \cite{Kirilyuk:05}. 


The above three techniques potentially measure the faceted magnetization, but to further measure the faceted compressibility, a means must be found to vary the chemical potential, e.g., by doping \cite{chen2010massive}, or gating.


\section*{Acknowledgements}
P.Z and T.L.H. were supported by ARO MURI W911NF2020166.
A.A. was supported  by the Gordon and Betty Moore Foundation EPiQS Initiative through Grant No. GBMF 4305 and GBMF 8691 at the University of Illinois.

\appendix
\begin{widetext}
\section{Derivation of Eq.\eqref{eq: Mghf2} from local orbital magnetization}
\label{app: deri_maghf2}

\subsection{Derivation of Eq.\eqref{eq: Mghf2} under OBC}
To derive Eq.\eqref{eq: Mghf2} from Eq.\eqref{eq: mag_bulk} under OBC in a slab geometry where $z$ direction is open and $x$ and $y$ directions are periodic, we introduce $\ket{\bm{k}_{\perp},R_{z},a}=\sum_{\bm{R}_{\perp}}\sqrt{A_{xy, \mathrm{cell}}}e^{i\bm{k}_{\perp}\cdot \bm{R}_{\perp}}\ket{\bm{R},a}$, where $A_{xy, \mathrm{cell}}$ is the area in the $x$-$y$ plane of a primitive unit cell, and $\bm{k}_{\perp}=(k_{x},k_{y})$, and $\bm{R}_{\perp}=(R_{x},R_{y})$. The normalization of $\ket{\bm{k}_{\perp},R_{z},a}$ is $\inp{\bm{k}_{\perp},R_{z},a}{\bm{k}_{\perp}^{\prime},R^{\prime}_{z},b}=(2\pi)^{2}\delta^{(2)}(\bm{k}_{\perp}-\bm{k}^{\prime}_{\perp})\delta_{R_{z}R_{z}^{\prime}}\delta_{ab}$, where $\delta^{(2)}(\bm{k}_{\perp}-\bm{k}^{\prime}_{\perp})$ is the two-dimensional Dirac delta function. Note that
\begin{equation}
\label{eq: baro_tildeo}
\bra{\bm{k}_{\perp},R_{z},a}\bar{O}\ket{\bm{k}_{\perp}^{\prime},R_{z}^{\prime},b}=(2\pi)^2\delta^{(2)}(\bm{k}_{\perp}-\bm{k}^{\prime}_{\perp})\widetilde{O}_{R_{z}R_{z}^{\prime},ab},  \quad \mathrm{for} \quad O=P,Q,H.
\end{equation}
Insert the identity $(1/(2\pi)^2)\sum_{R_{z},a}\int_{rBZ}d\bm{k}_{\perp}\ket{\bm{k}_{\perp},R_{z},a}\bra{\bm{k}_{\perp},R_{z},a}=\mathbbm{1}$ into Eq.\eqref{eq: mag_bulk}, and sum over $\bm{R}_{\perp}$ fixing $R_{z}=l$, we have
\begin{equation}
\label{eq: localmag_layermag}
\begin{aligned}
&\frac{1}{A}\sum_{\bm{R}_{\perp}}\mathfrak{m}_{z}(\bm{R}_{\perp},R_{z}=l)=\frac{1}{A}\sum_{\bm{R}_{\perp}}\frac{e}{\hbar c}\operatorname{Im}\sum_{a,b,c}\sum_{R_{z}^{\prime},R_{z}^{\prime\prime}}\int_{rBZ}\frac{d\bm{k}_{\perp}}{(2\pi)^{2}}\frac{d\bm{k}^{\prime}_{\perp}}{(2\pi)^{2}}\inp{\bm{R}_{\perp},l,a}{\bm{k}_{\perp},R_{z}^{\prime},b}\bra{\bm{k}_{\perp},R_{z}^{\prime},b}
\\
&\times\bar{P} \hat{r}_{x}\bar{Q} \bar{H} \bar{Q} \hat{r}_{y} \bar{P}-\bar{Q}\hat{r}_{x}\bar{P}(\bar{H}-2\mu)\bar{P}\hat{r}_{y}\bar{Q}\ket{\bm{k}^{\prime}_{\perp},R_{z}^{\prime\prime},c}\inp{\bm{k}^{\prime}_{\perp},R_{z}^{\prime\prime},c}{\bm{R}_{\perp},l,a}
\\
&=\frac{1}{A}\sum_{\bm{R}_{\perp}}A_{xy,\mathrm{cell}}\frac{e}{\hbar c}\operatorname{Im}\int_{rBZ}\frac{d\bm{k}_{\perp}}{(2\pi)^{2}}\frac{d\bm{k}^{\prime}_{\perp}}{(2\pi)^{2}}e^{i(\bm{k}_{\perp}-\bm{k}^{\prime}_{\perp})\cdot \bm{R}_{\perp}}\bra{\bm{k}_{\perp},l,a}\bar{P} \hat{r}_{x}\bar{Q} \bar{H} \bar{Q} \hat{r}_{y} \bar{P}-\bar{Q}\hat{r}_{x}\bar{P}(\bar{H}-2\mu)\bar{P}\hat{r}_{y}\bar{Q}\ket{\bm{k}^{\prime}_{\perp},l,a},
\end{aligned}
\end{equation}
where $A$ is the area of the $x$-$y$ plane of the system, and $A=\sum_{\bm{R}_{\perp}}A_{xy,\mathrm{cell}}$. With Eq.\eqref{eq: baro_tildeo}, follow steps closely analogous to Eq.\eqref{eq: g2}, we can derive
\begin{equation}
\label{eq: g3}
\begin{aligned}
&\sum_{a}\bra{\bm{k}_{\perp},l,a}\bar{P} \hat{r}_{x}\bar{Q} \bar{H} \bar{Q} \hat{r}_{y} \bar{P}\ket{\bm{k}^{\prime}_{\perp},l,a}
\\
&=(2\pi)^{2}\delta^{(2)}(\bm{k}_{\perp}-\bm{k}^{\prime}_{\perp})\sum_{R_{z}^{1},R_{z}^{2},R_{z}^{3},R_{z}^{4}}\operatorname{Tr}_{\mathrm{cell}}\left[\partial_{k_{x}}\widetilde{P}_{lR_{z}^{1}}\widetilde{Q}_{R_{z}^{1}R_{z}^{2}}\widetilde{H}_{R_{z}^{2}R_{z}^{3}}\widetilde{Q}_{R_{z}^{3}R_{z}^{4}}\partial_{k_{y}}\widetilde{P}_{R_{z}^{4}l}\right]
\\
&=(2\pi)^{2}\delta^{(2)}(\bm{k}_{\perp}-\bm{k}^{\prime}_{\perp}) \operatorname{Tr}_{\mathrm{cell},z}\left[\mathcal{P}_{l}\partial_{k_{x}}\widetilde{P}\widetilde{Q}\widetilde{H}\widetilde{Q}\partial_{k_{y}}\widetilde{P}\right]=(2\pi)^{2}\delta^{(2)}(\bm{k}_{\perp}-\bm{k}^{\prime}_{\perp})g_{\bm{k}_{\perp},xy}(l),
\end{aligned}
\end{equation}
where $\mathcal{P}_{l}$ is the projector on the layer with $R_{z}=l$, and we remind that $\operatorname{Tr}_{\mathrm{cell},z}$ is the trace respective to the index $(R_{z},a)$. The last equality used the definition of $g_{\bm{k}_{\perp},xy}(l)$ in Eq.\eqref{eq: g&h2}. Similarly, we can also derive
\begin{equation}
\label{eq: h}
\begin{aligned}
&\bra{\bm{k}_{\perp},l,a}\bar{Q} \hat{r}_{x}\bar{P} \bar{H} \bar{P} \hat{r}_{y} \bar{Q}\ket{\bm{k}^{\prime}_{\perp},l,a}=-(2\pi)^{2}\delta^{(2)}(\bm{k}_{\perp}-\bm{k}^{\prime}_{\perp}) \operatorname{Tr}_{\mathrm{cell},z}\left[\mathcal{P}_{l}\widetilde{Q}\partial_{k_{x}}\widetilde{P}\widetilde{H}\widetilde{P}\partial_{k_{y}}\widetilde{Q}\right]
\\
&=(2\pi)^{2}\delta^{(2)}(\bm{k}_{\perp}-\bm{k}^{\prime}_{\perp}) \operatorname{Tr}_{\mathrm{cell},z}\left[\mathcal{P}_{l}\widetilde{Q}\partial_{k_{x}}\widetilde{P}\widetilde{H}\partial_{k_{y}}\widetilde{P}\widetilde{Q}\right]=(2\pi)^{2}\delta^{(2)}(\bm{k}_{\perp}-\bm{k}^{\prime}_{\perp})h_{\bm{k}_{\perp},yx}(l),
\end{aligned}
\end{equation}
and 
\begin{equation}
\label{eq: f}
\begin{aligned}
&\bra{\bm{k}_{\perp},l,a}\bar{Q} \hat{r}_{x}\bar{P}\bar{P} \hat{r}_{y} \bar{Q}\ket{\bm{k}^{\prime}_{\perp},l,a}=-(2\pi)^{2}\delta^{(2)}(\bm{k}_{\perp}-\bm{k}^{\prime}_{\perp}) \operatorname{Tr}_{\mathrm{cell},z}\left[\mathcal{P}_{l}\widetilde{Q}\partial_{k_{x}}\widetilde{P}P\partial_{k_{y}}\widetilde{Q}\right]
\\
&=(2\pi)^{2}\delta^{(2)}(\bm{k}_{\perp}-\bm{k}^{\prime}_{\perp}) \operatorname{Tr}_{\mathrm{cell},z}\left[\mathcal{P}_{l}\widetilde{Q}\partial_{k_{x}}\widetilde{P}\partial_{k_{y}}\widetilde{P}\widetilde{Q}\right]=(2\pi)^{2}\delta^{(2)}(\bm{k}_{\perp}-\bm{k}^{\prime}_{\perp})f_{\bm{k}_{\perp},yx}(l).
\end{aligned}
\end{equation}
In Eq.\eqref{eq: h} and Eq.\eqref{eq: f}, we used $\widetilde{P}\partial\widetilde{Q}=-\partial\widetilde{P}\widetilde{Q}$; the last steps of both Eq.\eqref{eq: h} and Eq.\eqref{eq: f} again used the definition of  $h_{\bm{k}_{\perp},yx}(l)$ and $f_{\bm{k}_{\perp},yx}(l)$ in Eq.\eqref{eq: g&h2} . Note that from  Eq.\eqref{eq: g&h2}, it is straightforward to see $\operatorname{Im}(\widetilde{o}_{\bm{k}_{\perp},xy})=-\operatorname{Im}(\widetilde{o}_{\bm{k}_{\perp},yx})$ for $o=g,h,f$. With this fact, we substitute Eq.\eqref{eq: g3}, Eq.\eqref{eq: h} and Eq.\eqref{eq: f} into Eq.\eqref{eq: localmag_layermag}, we can finally derive Eq.\eqref{eq: Mghf2}.

\subsection{Equivalence of Eq.\eqref{eq: mag_bulk} and Eq.\eqref{eq: Mghf} under PBC }

To prove Eq.\eqref{eq: mag_bulk} and Eq  .\eqref{eq: Mghf} with $\gamma=z$ are equivalent under PBC,  we define $\ket{\bm{k},a}$ as the Fourier transformation of $\ket{\bm{R},a}$ such that $\ket{\bm{k},a}=\sum_{\bm{R}}\sqrt{V_{\mathrm{cell}}}e^{i\bm{k}\cdot\bm{R}}\ket{\bm{R},a}$, where $V_{\mathrm{cell}}$ is the volume of a primitive unit cell and satisfies
$\sum_{\bm{R}}V_{\mathrm{cell}}=V$. Then normalization of $\ket{\bm{k},a}$ is $\inp{\bm{k},a}{\bm{k}^{\prime},b}=(2\pi)^{3}\delta^{(3)}(\bm{k}-\bm{k}^{\prime})\delta_{ab}$, where $\delta^{(3)}(\bm{k}-\bm{k}^{\prime})$ is the three-dimensional Dirac delta function. Note that
\begin{equation}
\label{eq: baro_o}
\bra{\bm{k},a}\bar{O}\ket{\bm{k}^{\prime},b}=(2\pi)^{3}\delta^{(3)}(\bm{k}-\bm{k}^{\prime})O_{ab}(\bm{k}), \quad \mathrm{for} \quad O=P,Q,H.
\end{equation}
We will show that 
\begin{equation}
\label{eq: g1}
\begin{aligned}
(2\pi)^{3}\delta^{(3)}(\bm{k}-\bm{k}^{\prime})g_{\mathbf{k},xy}=(2\pi)^{3}\delta^{(3)}(\bm{k}-\bm{k}^{\prime})\operatorname{Tr}_{\mathrm{cell}}\left[\partial_{k_{x}} PQHQ\partial_{k_{y}} P\right]=\sum_{a}\bra{\bm{k},a}\bar{P}\hat{r}_{x}\bar{Q}\bar{H}\bar{Q}\hat{r}_{y} \bar{P}\ket{\bm{k}^{\prime},a}.
\end{aligned}
\end{equation}
We remind that $\operatorname{Tr}_{\mathrm{cell}}$ is the trace respective to the unit-cell index $a$. We prove the above equation by inserting the identity $(1/(2\pi)^{3})\sum_{a}\int_{BZ}d\bm{k}\ket{\bm{k},a}\bra{\bm{k},a}=\mathbbm{1}$ into the RHS of the above equation: 
\begin{equation}
\label{eq: g2}
\begin{aligned}
&\sum_{a}\bra{\bm{k},a}\bar{P}\hat{r}_{x}\bar{Q}\bar{H}\bar{Q}\hat{r}_{y} \bar{P}\ket{\bm{k}^{\prime},a}=\int_{BZ} \frac{d\mathbf{k}^{1}}{(2\pi)^{3}}\frac{d\mathbf{k}^{3}}{(2\pi)^{3}}\frac{d\mathbf{k}^{3}}{(2\pi)^{3}}\frac{d\mathbf{k}^{4}}{(2\pi)^{3}}\sum_{a,b,c,d,e}\bra{\bm{k},a}\bar{P}\ket{\bm{k}^1,b}\bra{\bm{k}^1,b}\hat{r}_{x}\bar{Q}\ket{\bm{k}^2,c}\bra{\bm{k}^2,c}
\\
&\quad \times\bar{H}\ket{\bm{k}^3,d}\bra{\bm{k}^3,d}\bar{Q}\ket{\bm{k}^4,e}\bra{\bm{k}^4,e}\hat{r}_{y} \bar{P}\ket{\bm{k}^{\prime},a}
\\
&=(2\pi)^{3}\int_{BZ} d\mathbf{k}^{1}d\mathbf{k}^{2}d\mathbf{k}^{3}d\mathbf{k}^{4}\operatorname{Tr}_{\mathrm{cell}}\big\{P(\bm{k})\delta^{(3)}(\bm{k}-\bm{k}^{1}) i\partial_{k^{1}_{x}}\left[Q(\bm{k}^1)\delta^{(3)}(\bm{k}^1-\bm{k}^2)\right]
H(\bm{k}^2)\delta^{(3)}(\bm{k}^2-\bm{k}^3)Q(\bm{k}^3)
\\
&\quad\times \delta^{(3)}(\bm{k}^3-\bm{k}^4)i\partial_{k^4_{y}} \left[P(\bm{k}^4)\delta^{(3)}(\bm{k}^4-\bm{k}^{\prime})\right]\big\}
\\
&=(2\pi)^{3}\int_{BZ} d\mathbf{k}^{1}d\mathbf{k}^{2}d\mathbf{k}^{3}d\mathbf{k}^{4}\delta^{(3)}(\bm{k}-\bm{k}^1)\delta^{(3)}(\bm{k}^{1}-\bm{k}^2)\delta^{(3)}(\bm{k}^2-\bm{k}^3)\delta^{(3)}(\bm{k}^3-\bm{k}^4)\delta^{(3)}(\bm{k}^4-\bm{k}^{\prime})
\\
&\quad\times\operatorname{Tr}_{\mathrm{cell}}\left[P(\bm{k})i\partial_{k^{1}_{x}}Q(\bm{k}^1)H(\bm{k}^2)Q(\bm{k}^3)i\partial_{k^{4}_{y}} P(\bm{k}^4)\right]
\\
&=-(2\pi)^{3}\delta^{(3)}(\bm{k}-\bm{k}^{\prime})\operatorname{Tr}_{\mathrm{cell}}\left[ P(\bm{k})\partial_{k_{x}}Q(\bm{k})H(\bm{k})Q(\bm{k})\partial_{k_{y}} P(\bm{k})\right]
\\
&=(2\pi)^{3}\delta^{(3)}(\bm{k}-\bm{k}^{\prime})\operatorname{Tr}_{\mathrm{cell}}\left[ \partial_{k_{x}}P(\bm{k})Q(\bm{k})H(\bm{k})Q(\bm{k})\partial_{k_{y}} P(\bm{k})\right]=(2\pi)^{3}\delta^{(3)}(\bm{k}-\bm{k}^{\prime})g_{\bm{k},xy}
\end{aligned}
\end{equation}
From the second equality to the third equality in the above equation, we used the fact that $P(\bm{k})\delta^{(3)}(\bm{k}-\bm{k}^{1})Q(\bm{k}^1)=0$, so that we have 
\begin{equation}
\begin{aligned}
&P(\bm{k})\delta^{(3)}(\bm{k}-\bm{k}^{1})i\partial_{k^{1}_{x}}\left[Q(\bm{k}^1)\delta^{(3)}(\bm{k}^1-\bm{k}^2)\right]
\\
&=P(\bm{k})\delta^{(3)}(\bm{k}-\bm{k}^{1})\left\{\left[i\partial_{k^{1}_{x}}Q(\bm{k}^1)\right]\delta^{(3)}(\bm{k}^1-\bm{k}^{2})+Q(\bm{k}^1)\left[i\partial_{k^{1}_{x}}\delta^{(3)}(\bm{k}^1-\bm{k}^2)\right]\right\}
\\
&=P(\bm{k})\delta_{k_{z},k_{z}^{1}}\delta^{(3)}(\bm{k}-\bm{k}^{1})\left[i\partial_{k^{1}_{x}}Q(\bm{k}_1)\right]\delta^{(3)}(\bm{k}^1-\bm{k}^2),
\end{aligned}
\end{equation}
and 
\begin{equation}
\begin{aligned}
&Q(\bm{k}^{3})\delta^{(3)}(\bm{k}^3-\bm{k}^4)i\partial_{k^{4}_{y}}\left[P(\bm{k}^4)\delta^{(3)}(\bm{k}^4-\bm{k}^{\prime})\right]
\\
&=Q(\bm{k}^3)\delta^{(3)}(\bm{k}^3-\bm{k}^{4})\left\{\left[i\partial_{k^{4}_{y}}P(\bm{k}^4)\right]\delta^{(3)}(\bm{k}^4-\bm{k}^{\prime})+P(\bm{k}^4)\left[i\partial_{k^{4}_{y}}\delta^{(3)}(\bm{k}^4-\bm{k}^{\prime})\right]\right\}
\\
&=Q(\bm{k}^3)\delta^{(3)}(\bm{k}^3-\bm{k}^{4})\left[i\partial_{k^{4}_{y}}P(\bm{k}^4)\right]\delta^{(3)}(\bm{k}^4-\bm{k}^{\prime}).
\end{aligned}
\end{equation}
Following the same procedure, we can also get 
\begin{equation}
\label{eq: hf1}
\begin{array}{c}
(2\pi)^{3}\delta^{(3)}(\bm{k}-\bm{k}^{\prime})h_{\mathbf{k},yx}=\sum_{a}\bra{\bm{k},a}\bar{Q}\hat{r}_{x}\bar{P}\bar{H}\bar{P}\hat{r}_{y} \bar{Q}\ket{\bm{k}^{\prime},a},
\\
(2\pi)^{3}\delta^{(3)}(\bm{k}-\bm{k}^{\prime})f_{\mathbf{k},yx}=\sum_{a}\bra{\bm{k},a}\bar{Q}\hat{r}_{x}\bar{P}\bar{P}\hat{r}_{y} \bar{Q}\ket{\bm{k}^{\prime},a}.
\end{array}
\end{equation}
Then, by inserting the identity $(1/(2\pi)^{3})\sum_{a}\int_{BZ}d\bm{k}\ket{\bm{k},a}\bra{\bm{k},a}=\mathbbm{1}$ into Eq.\eqref{eq: mag_bulk}, we have
\begin{equation}
\label{eq: localmag_ksapce2}
\begin{aligned}
&M_{z}=\frac{1}{V}\sum_{\bm{R}}\mathfrak{m}_{z}(\bm{R})=\frac{1}{V}\sum_{\bm{R}}\frac{e}{\hbar c}\operatorname{Im}\sum_{a,b,c}\int_{BZ}\frac{d\bm{k}}{(2\pi)^{3}}\frac{d\bm{k}^{\prime}}{(2\pi)^{3}}\inp{\bm{R},a}{\bm{k},b}\bra{\bm{k},b}\bar{P} \hat{r}_{x}\bar{Q} \bar{H} \bar{Q} \hat{r}_{y} \bar{P}
\\
&\quad-\bar{Q}\hat{r}_{x}\bar{P}(\bar{H}-2\mu)\bar{P}\hat{r}_{y}\bar{Q}\ket{\bm{k}^{\prime},c}\inp{\bm{k}^{\prime},c}{\bm{R},a}
\\
&=\frac{1}{V}\sum_{\bm{R}}\frac{e}{\hbar c}\operatorname{Im}\sum_{a,b,c}V_{\mathrm{cell}}\delta_{ab}\delta_{ca}\int_{BZ}\frac{d\bm{k}}{(2\pi)^{3}}\frac{d\bm{k}^{\prime}}{(2\pi)^{3}}e^{i(\bm{k}-\bm{k}^{\prime})\cdot\bm{R}}\bra{\bm{k},b}\bar{P} \hat{r}_{x}\bar{Q} \bar{H} \bar{Q} \hat{r}_{y} \bar{P}-\bar{Q}\hat{r}_{x}\bar{P}(\bar{H}-2\mu)\bar{P}\hat{r}_{y}\bar{Q}\ket{\bm{k}^{\prime},c}
\\
&=\frac{1}{V}\sum_{\bm{R}}\frac{e}{\hbar c}\operatorname{Im}\sum_{a}V_{\mathrm{cell}}\int_{BZ}\frac{d\bm{k}}{(2\pi)^{3}}\frac{d\bm{k}^{\prime}}{(2\pi)^{3}}e^{i(\bm{k}-\bm{k}^{\prime})\cdot\bm{R}}\bra{\bm{k},a}\bar{P} \hat{r}_{x}\bar{Q} \bar{H} \bar{Q} \hat{r}_{y} \bar{P}-\bar{Q}\hat{r}_{x}\bar{P}(\bar{H}-2\mu)\bar{P}\hat{r}_{y}\bar{Q}\ket{\bm{k}^{\prime},a}
\\
&=\frac{1}{V}\sum_{\bm{R}}\frac{e}{\hbar c}\operatorname{Im}V_{\mathrm{cell}}\int_{BZ}\frac{d\bm{k}}{(2\pi)^{3}}d\bm{k}^{\prime}e^{i(\bm{k}-\bm{k}^{\prime})\cdot\bm{R}}\delta^{(3)}(\bm{k}-\bm{k}^{\prime})\left(g_{\bm{k},xy}-h_{\bm{k},yx}+2\mu f_{\bm{k},yx}\right)
\\
&=\frac{e}{\hbar c}\frac{1}{V}\sum_{\bm{R}}V_{\mathrm{cell}}\operatorname{Im}\int_{BZ}\frac{d\bm{k}}{(2\pi)^{3}}\left(g_{\bm{k},xy}-h_{\bm{k},yx}+2\mu f_{\bm{k},yx}\right)
\\
&=\frac{e}{\hbar c}\operatorname{Im}\int_{BZ}\frac{d\bm{k}}{(2\pi)^{3}}\left(g_{\bm{k},xy}-h_{\bm{k},yx}+2\mu f_{\bm{k},yx}\right)
\end{aligned}
\end{equation}
Note that from Eq.\eqref{eq: g&h}, it is straightforward to see $o_{\bm{k},xy}=o^{\star}_{\bm{k},yx}$ for $o=g,h,f$, which leads to $\operatorname{Im}(o_{\bm{k},xy})=-\operatorname{Im}(o^{\star}_{\bm{k},yx})$. With this fact, the above equation becomes
\begin{equation}
\label{eq: localmag_ksapce3}
\begin{aligned}
M_{z}=\frac{1}{V}\sum_{\bm{R}}\mathfrak{m}_{z}(\bm{R})
=\frac{e}{\hbar c}\operatorname{Im}\int_{BZ}\frac{d\bm{k}}{(2\pi)^{3}}\left(g_{\bm{k},xy}+h_{\bm{k},xy}-2\mu f_{\bm{k},xy}\right).
\end{aligned}
\end{equation}
If we fix $\gamma=z$ and $d=3$ in Eq.\eqref{eq: Mghf}, we have  
\begin{equation}
\begin{aligned}
M_{z}&=\frac{e}{c\hbar} \operatorname{Im}\int_{BZ} \frac{d \mathbf{k}}{(2 \pi)^{3}}\left\{\epsilon_{zxy}\bigg[\frac{1}{2}(g_{\mathbf{k},xy}+h_{\mathbf{k},xy})-\mu f_{\mathbf{k},xy}\bigg]+\epsilon_{zyx}\bigg[\frac{1}{2}(g_{\mathbf{k},yx}+h_{\mathbf{k},yx})-\mu f_{\mathbf{k},yx}\bigg]\right\}
\\
&=\frac{e}{c\hbar} \operatorname{Im}\int_{BZ} \frac{d \mathbf{k}}{(2 \pi)^{3}}\left\{\bigg[\frac{1}{2}(g_{\mathbf{k},xy}+h_{\mathbf{k},xy})-\mu f_{\mathbf{k},xy}\bigg]-\bigg[\frac{1}{2}(g_{\mathbf{k},yx}+h_{\mathbf{k},yx})-\mu f_{\mathbf{k},yx}\bigg]\right\}
\\
&=\frac{e}{c\hbar} \operatorname{Im}\int_{BZ} \frac{d \mathbf{k}}{(2 \pi)^{3}}(g_{\mathbf{k},xy}+h_{\mathbf{k},xy}-2\mu f_{\mathbf{k},xy}),
\end{aligned}
\end{equation}
which is clearly equivalent to Eq.\eqref{eq: localmag_ksapce3}.


\section{Transformation of the magnetization under crystallographic spacetime group}\label{app: Mtransform}

If the bulk Chern class is trivial, the bulk orbital magnetization has vanishing compressibility, and is simply related to a trace of the orbital angular momentum operator:
$\mathbf{M}=(-e/2Vc)\textit{tr}(\bar{P}\hat{\bm{r}}\times\hat{\bm{v}})$. Note``$\textit{tr}$" is the trace over the whole Hilbert space, $V$ is the volume of the system, and $\bar{P}$ is the projector onto the occupied subspace of a Hamiltonian $\bar{H}$ in the real space basis. The $\hat{\bm{r}}$ and $\hat{\bm{v}}$ are position operator and velocity operator respectively.

We show here that $\mathbf{M}$ transforms as a pseudo-vector under the crystallographic spatial transformations and is moreover odd under  time reversal.   Consider an element $g$ in the spacetime symmetry group such that
\begin{equation}
\label{eq: r_trans}
\hat{g}\hat{r}_{i}\hat{g}^{-1}=\breve{g}_{ij} \hat{r}_{j}+\delta_{i}, \breve{g}^{-1}=\breve{g}^{t} \in \mathbb{R}.
\end{equation}
If a Hamiltonian has the symmetry $\hat{g}$, $i.e.$, $\hat{g}\bar{H}\hat{g}^{-1}=\bar{H}$ and thus $\hat{g}\bar{P}\hat{g}^{-1}=\bar{P}$, given $\hat{v}_{l}=i[\bar{H},\hat{r}_{l}]/\hbar$, then we have
\begin{equation}
\label{eq: v_trans}
\begin{aligned}
\hat{g}\hat{v}_{l}\hat{g}^{-1}&=(-1)^{s(g)}(\breve{g}_{lj}\frac{i}{\hbar}[\bar{H},\hat{r}_{j}])
\\
&=(-1)^{s(g)}(\breve{g}_{lj}\hat{v}_{j}),
\end{aligned}
\end{equation}
where $s(g)=1$ if there is time-reversal in $g$, and otherwise $s(g)=0$. Substitute Eq.\eqref{eq: r_trans}
and Eq.\eqref{eq: v_trans} into $\mathbf{M}=(-e/2Vc)\textit{tr}(\bar{P}\hat{\bm{r}}\times\hat{\bm{v}})$, and note that $\textit{tr}(\bar{P}\hat{v}_{l})=0$ because the total current is zero in insulators under zero external field, we have
\begin{equation}
\begin{aligned}
\label{eq: m_trans1}
\mathbf{M}_{i}&=\frac{e}{2Vc}\epsilon_{ijl}\textit{tr}(\hat{g}\bar{P}\hat{g}^{-1}\hat{g}\hat{r}_{j}\hat{g}^{-1}\hat{g}\hat{v}_{l}\hat{g}^{-1})
\\
&=-\frac{e}{2Vc}(-1)^{s(g)}\epsilon_{ijl}\breve{g}_{jm}\breve{g}_{ln}\textit{tr}(\bar{P}\hat{r}_{m}\hat{v}_{n}).
\end{aligned}
\end{equation}
Multiply $\delta_{ip}=\breve{g}_{iq}\breve{g}^{-1}_{qp}=\breve{g}_{iq}\breve{g}_{pq}$ on both side of the above equation, given $\epsilon_{ijl}\breve{g}_{iq}\breve{g}_{jm}\breve{g}_{ln}=\det{\breve{g}}\epsilon_{qmn}$, we can finally derive
\begin{equation}
\label{eq: m_trans2}
\mathbf{M}=(-1)^{s(g)}\det{\breve{g}}(\breve{g}\mathbf{M}).
\end{equation}
Note $\det{\breve{g}}=-1$ for improper rotations, hence $\mathbf{M}$ is invariant under spatial inversion.


\section{Vanishing bulk orbital magnetization due to the antisymmetry}
\label{app: proof_hb}

To prove that the bulk orbital magnetization can be ruled out by the antisymmetry in Eq. \eqref{eq:antisymmetry}, let us first rewrite the orbital magnetization in the momentum representation:\cite{ceresoli2006orbital,shi2007quantum}
\begin{equation}
\label{eq: Mghf}
\begin{aligned}
M_{\gamma}&=\frac{e}{c\hbar} \operatorname{Im}\int_{BZ} \frac{d \mathbf{k}}{(2 \pi)^{d}}\epsilon_{\alpha\beta\gamma}\bigg[\frac{1}{2}(g_{\mathbf{k},\alpha\beta}+h_{\mathbf{k},\alpha\beta})-\mu f_{\mathbf{k},\alpha\beta}\bigg],
\end{aligned}
\end{equation}
where
\begin{equation}
\label{eq: g&h}
\begin{array}{c}g_{\mathbf{k}, \alpha\beta}=\operatorname{Tr}_{\mathrm{cell}}\left[\partial_{\alpha} P QHQ\partial_{\beta} P\right],
\\ 
h_{\mathbf{k}, \alpha \beta}=\operatorname{Tr}_{\mathrm{cell}}\left[ Q\partial_{\beta} PH\partial_{\alpha} PQ\right],
\\
f_{\mathbf{k}, \alpha \beta}=\operatorname{Tr}_{\mathrm{cell}}\left[ Q\partial_{\beta} P\partial_{\alpha} PQ\right],
\end{array}
\end{equation}
$H(\mathbf{k})$ is a Bloch Hamiltonian with an energy gap separating occupied and unoccupied bands, which are respectively projected by $P(\bk)$ and $Q(\bk)$. (In the above equation, the arguments of $P, Q, H$ have been omitted for simplicity.)  All of $P(\bk),Q(\bk)$ and $H(\bk)$ are matrices with row index $a$ labelling the basis vectors of the Hilbert space of one primitive unit cell, and ``$\operatorname{Tr}_{\mathrm{cell}}$" is the trace with respect to the unit-cell index $a$. If  $H(\mathbf{k})$ is an insulating Bloch Hamiltonian satisfying \q{generalantisymmetry}, then
$U P^{\star}(\mathbf{k})U^{\dag}=Q(\mathbf{k})$. Substituting this into Eq.\eqref{eq: g&h}, we have
\begin{equation}
\begin{aligned}
&h_{\mathbf{k}, \alpha \beta}=\operatorname{Tr}_{\mathrm{cell}}\left[\partial_{\beta} PH\partial_{\alpha} P Q\right]=\operatorname{Tr}_{\mathrm{cell}}\left[\partial_{\beta} PH\partial_{\alpha} PU P^{\star}U^{\dag}\right]
\\
&=\operatorname{Tr}_{\mathrm{cell}}[\partial_{\beta} PUU^{\dag}HUU^{\dag}\partial_{\alpha} PU P^{\star}U^{\dag}]
\\
&=\operatorname{Tr}_{\mathrm{cell}}[(U^{\dag}\partial_{\beta} PU)(U^{\dag}HU)(U^{\dag}\partial_{\alpha} PU) P^{\star}]
\\
&=-\operatorname{Tr}_{\mathrm{cell}}\left[\partial_{\beta} Q^{\star}H^{\star}\partial_{\alpha} Q^{\star} P^{\star}\right]
\\
&=-\operatorname{Tr}_{\mathrm{cell}}\left[\partial_{\beta} P^{\star}H^{\star}\partial_{\alpha} P^{\star} P^{\star}\right]
\\
&=-\operatorname{Tr}_{\mathrm{cell}}\left[P\partial_{\alpha} PH\partial_{\beta} P\right].
\end{aligned}
\end{equation}
For $g_{\mathbf{k},\alpha\beta}$ in Eq.\eqref{eq: g&h}, using $\partial_{\alpha}PQ=P\partial_{\alpha}P$ and $Q\partial_{\beta}P=\partial_{\beta}PP$ we can do some manipulations:
\begin{equation}
\begin{aligned}
&g_{\mathbf{k}, \alpha \beta}=\operatorname{Tr}_{\mathrm{cell}}\left[\partial_{\alpha} P QHQ\partial_{\beta} P\right]=\operatorname{Tr}_{\mathrm{cell}}\left[P\partial_{\alpha} PH\partial_{\beta} PP\right]
\\
&=\operatorname{Tr}_{\mathrm{cell}}\left[P\partial_{\alpha} PH\partial_{\beta} P\right].
\end{aligned}
\end{equation}
Compare the above two equations, we can clearly see $h_{\bm{k},\alpha\beta}=-g_{\bm{k},\alpha\beta}$. Thus, the first two terms of Eq.\eqref{eq: Mghf} cancel each other. The third and last term is proportional to the bulk-BZ-integral of $f_{\mathbf{k}, \alpha \beta}$, and therefore vanishes along with the bulk Chern invariants, completing our proof.

\end{widetext}
\bibliography{apssamp}

\end{document}